\documentstyle[amssymb,prd,aps,epsfig]{revtex}

\begin{document}
%
%
\def\be{\begin{equation}}
\def\ee{\end{equation}}
\def\RR{{\Bbb{R}}}
\def \wg#1{\mbox{\boldmath ${#1}$}}
\newfont{\ssg}{cmssbx10}
\def \w#1{\mbox{\ssg {#1}}}
\draft
\title{Binary black holes in circular orbits. II. Numerical methods and first results}
\author{Philippe Grandcl\'ement\footnote{present address: 
Department of Physics and Astronomy, Northwestern University, 
Evanston, IL 60208, USA;
e-mail: {\tt PGrandclement@northwestern.edu}}, 
Eric Gourgoulhon\footnote{e-mail: {\tt Eric.Gourgoulhon@obspm.fr}} 
and Silvano Bonazzola\footnote{e-mail: {\tt Silvano.Bonazzola@obspm.fr}} }
\address{D\'epartement d'Astrophysique Relativiste et de Cosmologie \\
  UMR 8629 du C.N.R.S., Observatoire de Paris, \\
  F-92195 Meudon Cedex, France
}
\date{11 October 2001}
\maketitle

\begin{abstract}
We present the first results from a new method for computing spacetimes representing
corotating binary black holes in circular orbits. The method is based on the assumption of
exact equilibrium. It uses the standard 3+1 
decomposition of Einstein equations  and conformal flatness approximation for the 
3-metric. Contrary to previous numerical approaches to this problem, 
we do not solve only the constraint equations but rather a set of five equations for 
the lapse function, the conformal factor and the
shift vector. The orbital velocity is unambiguously determined by
imposing that, at infinity, the metric behaves like the Schwarzschild one,
a requirement which is equivalent to the virial theorem. 
The numerical scheme has been implemented using multi-domain spectral methods and 
passed numerous tests.
A sequence of corotating black holes of equal mass is calculated.
Defining the sequence by requiring that the ADM mass decrease is equal
to the angular momentum decrease multiplied by the orbital angular velocity,
it is found that the area of the apparent horizons is constant along the sequence.
We also find a turning point in the ADM mass and angular
momentum curves, which may be interpreted as an innermost stable circular orbit 
(ISCO). The values of the global quantities at the ISCO, especially the orbital
velocity, are in much better agreement with those from third post-Newtonian
calculations than with those resulting from previous numerical approaches.
\end{abstract}


\begin{center}
We dedicate this work to the memory of our friend and collaborator 
Jean-Alain Marck.
\end{center}

\section{Introduction}\label{s:introduction}
Motivated by the construction of several gravitational wave detectors (LIGO, 
GEO600, TAMA300 and VIRGO) great efforts have been conducted in the past years 
to compute the waves generated by binary black holes. 
We presented in Ref. \cite{GourgGB01} 
(hereafter Paper I) a new method for getting quasi-stationary spacetimes 
representing binary black holes in circular orbits. See also 
Paper I for a review on issues and previous works in this field.

The basic approximation is to assume the existence of an helical Killing vector~:
\be
\wg{\ell} = \frac{\partial}{\partial t_0}+\Omega \frac{\partial}
{\partial \varphi_0},
\ee
where $\partial / \partial t_0$ (resp. $\partial / \partial \varphi_0$) is a timelike 
(resp. spacelike) vector which coincides asymptotically with the time coordinate
(resp. azimuthal coordinate) vector of an asymptotically inertial observer. Basically, it
means that the two black holes are on circular orbits with orbital velocity 
$\Omega$ \cite{Detwe89}. This is of course not exact because the 
emission of gravitational waves will cause the two holes 
to spiral toward each other. But this is a valid approximation as long as the 
time-scale of the gravitational radiation is much longer than the orbital 
period, which should be true, at least for large separations. The existence 
of $\wg{\ell}$ enables us to get rid of any time evolution.

We use the standard 3+1 decomposition of the Einstein equations \cite{York79}. We
 restrict ourselves to a space metric that is conformally flat,
i.e. of the form~:
\be
\wg{\gamma} = \Psi^4 \w{f},
\ee
where $\Psi$ is a scalar field and $\w{f}$ denotes the flat 3-metric
\cite{MatheMW98}. Let 
us mention that the exact spacetime should differ from conformal flatness
and that this assumption is only introduced for simplification and should be 
removed from later works. However it is important to note that it is consistent
with the existence of the helical Killing vector and the assumption of asymptotic flatness.
The ten Einstein equations 
then reduce to five equations, one for the lapse function $N$, one for the 
conformal factor $\Psi$ and three for the shift vector $\vec{\beta}$ (see Paper I for 
derivation)~:

\begin{eqnarray}
\label{e:eq_lapse}
\Delta N &=& N\Psi^4  \hat{A}_{ij} \hat{A}^{ij} -2\bar{D}_j\ln \Psi 
\bar{D}^jN\\
\label{e:eq_beta}
\Delta \beta^i +\frac{1}{3} \bar{D}^i \bar{D}_j \beta^j &=& 
2\hat{A}^{ij}\left(\bar{D}_j N-6N\bar{D}_j \ln\Psi\right)\\
\label{e:eq_psi}
\Delta \Psi &=& -\frac{\Psi^5}{8} \hat{A}_{ij} \hat{A}^{ij}
\end{eqnarray}
where $\bar{D}_i$ denotes covariant derivative associated with $\w{f}$ and
$\Delta := \bar{D}_k \bar{D}^k$ the ordinary Laplace operator. $\hat{A}^{ij}$ 
is the reduced extrinsic curvature tensor related to $K^{ij}$ by 
$\hat{A}^{ij} := \Psi^4 K^{ij}$ and given by
\be
\label{e:def_A}
\hat{A}^{ij} = \frac{1}{2N}\left(L\beta\right)^{ij},
\ee
$\left(L\beta\right)^{ij}$ denoting the conformal Killing operator applied to the
shift vector
\be
\left(L\beta\right)^{ij} := \bar{D}^i \beta^j + \bar{D}^j \beta^i
-\frac{2}{3}\bar{D}_k \beta^k f^{ij}.
\ee

Equations (\ref{e:eq_lapse}), (\ref{e:eq_beta}) and (\ref{e:eq_psi})
are a set of five strongly elliptic equations that are coupled. 
To solve such a system, we must impose boundary conditions.
To recover the Minkowski spacetime at 
spatial infinity, i.e. asymptotical flatness, the fields must have the 
following behaviors~:

\begin{eqnarray}
\label{e:lapse_infinity}
N \rightarrow 1 &{\rm \phantom{000}when\phantom{000}}& r\rightarrow \infty\\
\label{e:beta_infinity}
{\bf \vec{\beta}} \rightarrow \Omega \frac{\partial}{\partial \varphi_0}
&{\rm \phantom{000}when\phantom{000}}& r \rightarrow \infty\\
\label{e:psi_infinity}
\Psi \rightarrow 1 &{\rm \phantom{000}when\phantom{000}}&
 r\rightarrow \infty.
\end{eqnarray}

As we wish to obtain solutions representing two black holes and not Minkowski
spacetime, we must impose a non-trivial spacetime topology. In Paper~I, we
define the topology to be that of the real line $\RR$ times the 3-dimensional 
Misner-Lindquist manifold \cite{Misne63,Lindq63} ; this defines two throats, 
being two disjointed spheres $S_1$ and $S_2$ of 
radii $a_1$ and $a_2$, centered on points $\left(x_1,0,0\right)$ and 
$\left(x_2,0,0\right)$ (such that $\left|x_1-x_2\right| > a_1+a_2$). 
Following Misner \cite{Misne63}, Lindquist\cite{Lindq63}, Kulkarny {\it et al.}
\cite{KulkaSY83}, Cook {\it et al.} \cite{Cook91,CookCDKMO93,Cook94} and others 
\cite{PfeifTC00,DieneJKN00}, we demand that the two 
sheets of the Misner-Lindquist manifold are isometric. Moreover we choose the lapse 
function $N$ to be antisymmetric with respect to this isometry. We solve the
Einstein equations only for the ``upper'' sheet, i.e. only for the space exterior 
to the throats, with boundary conditions given by~:

\begin{eqnarray}
\label{e:bound_lapse}
\left. N \right|_{S_1} = 0 &{\rm \phantom{000} and \phantom{000}}&
\left. N \right|_{S_2} = 0\\
\label{e:bound_beta}
\left. {\bf \vec{\beta}} \right|_{S_1} = 0 &{\rm \phantom{000} and \phantom{000}}&
\left. {\bf \vec{\beta}} \right|_{S_2} = 0\\
\label{e:bound_psi}
\left.\left(\frac{\partial \Psi}{\partial r_1}+\frac{\Psi}{2r_1}\right)
\right|_{S_1} = 0 &{\rm \phantom{000} and \phantom{000}}&
\left.\left(\frac{\partial \Psi}{\partial r_2}+\frac{\Psi}{2r_2}\right)
\right|_{S_2} = 0,
\end{eqnarray}
where $r_1$ and $r_2$ are the radial coordinates associated with spheres $S_1$
and $S_2$. Equations (\ref{e:bound_lapse}) reflect the antisymmetry of 
the lapse function $N$.
The boundary conditions for the shift vector, given by Eqs.
(\ref{e:bound_beta}), represent two black holes in {\em corotation} (rotation 
synchronized with the orbital motion), 
which is the only case studied in this paper. Those boundary conditions should be 
easily changed to represent other states of rotation (like {\em irrotation}). Equations
(\ref{e:bound_psi}) come from the isometry solely.

The orbital velocity $\Omega$ only appears in the boundary condition for the 
shift (see Eq. (\ref{e:beta_infinity})). Equations (\ref{e:eq_lapse}), 
(\ref{e:eq_beta}) and (\ref{e:eq_psi}) can be solved for any value of
$\Omega$. So we need an extra condition to fix the right value for $\Omega$. 
This is done by imposing that, at spatial infinity, the metric behaves like 
a Schwarzschild metric, i.e. by imposing that $\Psi^2 N$ has no monopolar term in $1/r$~:
\be
\label{e:condition_omega}
\Psi^2 N \sim 1+\frac{\alpha}{r^2} {\mathrm \phantom{000}when \phantom{000}}
r\rightarrow \infty.
\ee
In other words, $\Omega$ is chosen so that the ADM and the ``Komar-like'' masses
coincides, those masses being given by
\begin{eqnarray}
M_{\rm ADM} &=& -\frac{1}{2\pi}\oint_{\infty} \bar{D}^i \Psi dS_i
					\label{e:m_adm} \\
M_{\rm Komar} &=& \frac{1}{4\pi}\oint_{\infty} \bar{D}^i N dS_i.
					\label{e:m_komar}
\end{eqnarray}
As shown in \cite{GourgB94} and in Paper~I this is closely
linked to the virial theorem for stationary spacetimes. We will see later that 
this uniquely determines the orbital velocity, and that this velocity
tends to the Keplerian one at large separation.

This paper is organized as follows. Sec. \ref{s:numerical} is dedicated to the
presentation of the numerical scheme, that is based on multi-domain spectral
methods. In Sec. \ref{s:tests} we present some tests passed by the code, 
which encompass comparison with the Schwarzschild and Kerr black hole and 
the Misner-Lindquist solution \cite{Misne63,Lindq63}.
In Sec. \ref{s:sequence} we present results about a sequence 
of binary black holes in circular orbits. In particular we locate the 
innermost stable circular orbit and compare its location with other works. 
Sec. \ref{s:conclusion} is concerned with extension of this
work, for getting more complicated and more realistic results.

\section{Numerical treatments}\label{s:numerical}
\subsection {Multi-domain spectral methods} \label{s:split}
The numerical treatments used to solve the elliptic equations 
presented above is based on the same methods that we already successfully applied to 
binary neutron stars \cite{GourgGTMB01}. The sources of the equations being 
mainly concentrated around each hole we use two sets of polar coordinates 
centered around each throat (see Sec. \ref{s:introduction}). 
Note however that 
the tensorial basis of decomposition is a Cartesian one. For example, a vector 
field $\vec{V}$ will be given by its components on the Cartesian basis 
$\left(V_x, V_y, V_z\right)$ but each component is a function of the polar 
coordinates $\left(r, \theta, \varphi\right)$ with respect to the center 
of one hole or the other.

We use spectral methods to solve the elliptic equations presented in Sec. 
\ref{s:introduction} ; the fields are given 
by their expansion onto some basis functions. Mainly, we use expansion on 
spherical harmonics with respect to the angles $\left(\theta, \varphi\right)$ and 
Chebyshev polynomials for the radial coordinate. Let us mention that there 
exists two equivalent descriptions : a function can be given in the 
{\it coefficient space}, i.e. by the coefficients of its spectral expansion, or 
in the {\it configuration space} by specifying its value at some collocation
points \cite{CanutHQZ88}.

The sources of the elliptic equations being non-compactly supported, we must use 
a computational domain extending to infinity. This is done by dividing space into 
several types of domains~:
\begin{itemize}
\item a {\it kernel}, a sphere containing the origin of the polar coordinates 
centered on one of the throats.
\item several spherical {\it shells} extending to finite radius.
\item a {\it compactified domain} extending to infinity by the use of the 
computational coordinate $u = \frac{1}{r}$.
\end{itemize}
This technique enables us to choose the basis function so that the fields 
are regular everywhere, especially on the rotation axis and to impose exact 
boundary conditions at infinity. This has been presented with more 
details elsewhere \cite{GourgGTMB01,BonazGM98,BonazGM99,GrandBGM01}.
Note that since the last domain extends to spacelike infinity, the
surface integrals defining global quantities, such as (\ref{e:m_adm}) and
(\ref{e:m_komar}), can be computed without any approximation. This
contrasts with other numerical methods based on finite domains
(cf. e.g. Ref.~\cite{JanseDKN01} and Fig.~1 therein).
As two different sets of
coordinates are used, one centered on each hole, we are left with two
computational domains of this type, each describing all space and so 
overlapping.

The sources of the equations being concentrated around the two throats, we wish 
to split the total equations (\ref{e:eq_psi}), (\ref{e:eq_beta}) and 
(\ref{e:eq_lapse}) into two parts, each being centered mainly around each hole and 
solved using the associated polar coordinates set. So an equation of the 
type $\Delta F = G$ will be split into
\begin{eqnarray}
\Delta F_1 &=& G_1 \\
\Delta F_2 &=& G_2, 
\end{eqnarray}
with $F=F_1+F_2$ and $G=G_1+G_2$.
$G_a$ is constructed to be mainly concentrated around hole $a$, and so well 
described by polar coordinates around this hole. Therefore, the solved 
equations are~:
\begin{eqnarray}
\label{e:split_lapse} \Delta N_a &=& N \Psi^4 \hat{A}_{ij} \hat{A}^{ij}_a 
-\frac{2}{\Psi} \bar{D}^j \Psi_a \bar{D}_j N\\
\label{e:split_beta} \Delta \beta^i_a +\frac{1}{3}\bar{D}^i\bar{D}_j\beta^j_a 
&=& 2\hat{A}^{ij}\left(\bar{D}_jN_a-6\frac{N}{\Psi}\bar{D}_j \Psi_a\right) \\
\label{e:split_psi} \Delta \Psi_a &=& -\frac{\Psi^5}{8} \hat{A}_{ij} 
\hat{A}^{ij}_a ,
\end{eqnarray}
where the values with no index represent the total values and the values 
with index $a$ represent the values ``mostly''
generated by hole $a$ ($a=1$ or $2$). 
For example, we have $\bar{D}_i N = \bar{D}_i N_1 + \bar{D}_i N_2$, 
$\bar{D}_iN_a$ being concentrated around hole $a$. Doing so, the physical 
equations and sources are given by the sum of equations (\ref{e:split_psi}), 
(\ref{e:split_beta}) and (\ref{e:split_lapse}) for $a=1$ and $a=2$. For more 
details about such a splitting of the equations into two parts we refer 
to \cite{GourgGTMB01}.

\subsection {Elliptic equations solvers}
\subsubsection {Scalar Poisson equation solver with boundary condition on 
a single throat}\label{s:scal_poisson}
Using spectral methods with spherical harmonics, the resolution of the 
scalar Poisson equation reduces to the inversion of banded matrices. We 
already presented in details in \cite{BonazGM99,GrandBGM01} the methods to 
solve such equations in all space, imposing regularity at the origin and 
exact boundary condition at infinity. In the case of black holes we wish 
to replace the regularity at the origin by boundary conditions on the 
spheres $S_1$ and $S_2$ and to solve only for the part of space exterior to 
those spheres. 
In Ref. \cite{GrandBGM01} we have shown that, for each couple of indices 
 $\left(l,m\right)$ of a particular spherical harmonic, we can calculate 
one particular solution in each domain, two homogeneous solutions in the 
shells and only one in the kernel (due to regularity) and one in the external
domain (due to boundary condition at infinity). The next step was to determine 
the coefficients of the homogeneous solutions by imposing that the global 
solution is ${\mathcal C}^1$ at the boundaries between the different domains.

In the case of a single throat $S$, the boundary condition is given 
by a function of the angles solely, i.e. $B\left(\theta, \varphi\right)$.
One wishes 
to impose that the solution or its radial derivative is equal to $B$ on the 
sphere which corresponds respectively to a Dirichlet or a Neumann problem.
We choose the kernel so that its spherical boundary coincides with the throat.
So we do not solve in the kernel with represents the 
interior of the sphere. $B$ is expanded in spherical harmonics and for each 
couple $\left(l,m\right)$, we use one of the homogeneous solution in the 
first shell to fulfill the Dirichlet or Neumann boundary condition. After 
that we are left with one particular solution in every domain, one 
homogeneous solution in the innermost shell and in the external domain and 
two in the other shells. The situation is exactly the same as when a 
solution was sought in all space and the coefficients of the remaining 
homogeneous solutions are chosen to maintain continuity of the solution
 and of its first derivative. So the generalization of the scheme presented 
in \cite{BonazGM99,GrandBGM01} is straightforward and enables us to 
solve either the Dirichlet or Neumann problem, with any boundary condition 
imposed on the throat.

\subsubsection {Vectorial Poisson equation solver with boundary condition on
 a single throat} \label{s:vect_poisson}
We presented extensively two different schemes to solve the vectorial Poisson
equation (\ref{e:eq_beta}) in all space in \cite{GrandBGM01} (the 
Oohara-Nakamura \cite{OoharN97} and Shibata \cite{OoharNS87} schemes). 
We present here
 an extension of the so-called Oohara-Nakamura scheme to impose boundary 
condition a throat and to solve only for the exterior part of space. The 
Shibata scheme has not been chosen because, the solution being constructed 
from auxiliary quantities, it is not obvious  at all to impose boundary 
conditions on it. This is not the case with the Oohara-Nakamura scheme 
where the final solution is calculated directly as the solution of three scalar 
Poisson equations. More precisely the solution of (cf. Eq. \ref{e:split_beta})
\be
\Delta \beta^i + \lambda \bar{D}^i \bar{D}_j \beta^j = V^i
\phantom{0000} \left(\lambda \not= -1\right)
\ee
is found by solving the set of three scalar Poisson equations 
\be
\label{e:part_beta_vec}
\Delta \beta^i = V^i -\lambda \bar{D}^i \chi,
\ee
where $\chi$ is solution of 
\be
\label{e:part_chi_vec}
\Delta \chi = \frac{1}{\lambda+1} \bar{D}_i V^i.
\ee
Let us mention that this scheme should only be used with a source $\vec{V}$ 
that is continuous.
We use the scalar Poisson equation solvers  with boundary condition 
 previously described to solve for each Cartesian component of (\ref{e:part_beta_vec}) 
with the appropriate 
boundary conditions. But let us recall (see \cite{GrandBGM01}) that the Oohara-Nakamura 
scheme is only applicable if 
\be
\label{e:val_chi}
\chi = \bar{D}_i \beta^i
\ee
 and that it only ensures that 
\be
\label{e:delta_chi}
\Delta \left(\chi-\bar{D}_i \beta^i\right) = 0.
\ee
One can easily show that (\ref{e:delta_chi}) implies (\ref{e:val_chi}) if and 
only if
\be
\label{e:bound_chi}
\left. \chi \right|_S = \left. \bar{D}_i \beta^i \right|_S,
\ee
which is the boundary condition we must impose during the resolution of 
(\ref{e:part_chi_vec}) to use this scheme. Let us mention that $\chi$ being 
calculated before $\vec{\beta}$, we must use some iterative procedure. We first
solve (\ref{e:part_chi_vec}) with an initial guess of the boundary condition
and then determine $\vec{\beta}$ by solving (\ref{e:part_beta_vec}). Using 
that value, we can determine a new boundary condition for $\chi$, using 
(\ref{e:bound_chi}), and so a 
new $\vec{\beta}$. This procedure is repeated until it has sufficiently 
converged. The 
obtained $\vec{\beta}$ is then solution of the vectorial Poisson equation 
with either a Dirichlet or Neumann type boundary condition on the sphere $S$.

\subsubsection{Elliptic solvers with boundary conditions on two throats}
In order to illustrate how boundary conditions are put on the two spheres
$S_1$ and $S_2$, let us
concentrate on the Dirichlet problem for the scalar Poisson equation. One 
wishes to solve
\be
\label{e:poisson_split_bound}
\Delta F = G,
\ee
with the boundary conditions
\begin{eqnarray}
\label{e:bound_tot_un}\left. F \right|_{S_1} &=& B_1\left(\theta_1, \varphi_1\right) \\
\label{e:bound_tot_deux}\left. F \right|_{S_2} &=& B_2\left(\theta_2, \varphi_2\right),
\end{eqnarray}
where $B_1$ and $B_2$ are arbitrary functions. As 
explained in Sec. \ref{s:split}, the total equation is split into two parts
\begin{eqnarray}
\label{e:part_un} \Delta F_1 &=& G_1 \\
\label{e:part_deux} \Delta F_2 &=& G_2,
\end{eqnarray}
the equation labeled $a=1$ or $2$, being solved on the grid centered around 
hole $a$ so that the sphere $S_a$ coincides with the 
innermost boundary of the first shell.

During the first step we solve Eqs. (\ref{e:part_un}) and (\ref{e:part_deux})
with the boundary conditions
\begin{eqnarray}
\label{e:bound_auto_un}\left. F_1 \right|_{S_1} &=& B_1 \\
\label{e:bound_auto_deux}\left. F_2 \right|_{S_2} &=& B_2
\end{eqnarray}
by means of the scalar Poisson equation solver described in Sec.
\ref{s:scal_poisson}. Doing so, the 
total solution $F= F_1+F_2$ does not fulfill the boundary conditions
(\ref{e:bound_tot_un})-(\ref{e:bound_tot_deux}). So we calculate the 
values of $F_1$ on the sphere $S_2$ and modify 
the boundary condition (\ref{e:bound_auto_deux}) by 
$B'_2 = B_2 - \left. F_1 \right|_{S_2}$. The same 
modification is done with the boundary condition (\ref{e:bound_auto_un}). 
Then we solve once again 
for $F_1$ and $F_2$. The whole procedure is repeated until a sufficient
convergence is achieved. So we 
are left with a function $F$ which is solution of the Poisson equation 
(\ref{e:poisson_split_bound}) and 
which fulfills a given Dirichlet-type boundary condition on two spheres 
(\ref{e:bound_tot_un})-(\ref{e:bound_tot_deux}).

The same thing can be done for the Neumann problem by modifying the 
boundary conditions using the radial derivatives of the functions $F_a$. The 
same technique is applied for the vectorial Poisson equation. Let us 
mention that the iteration on the boundary conditions for $\vec{\beta}$, resulting from
the presence of 
the two spheres, is done at the same time than the one on the quantity
$\chi$ resulting from to the Oohara-Nakamura scheme 
(see Sec. \ref{s:vect_poisson}).

\subsubsection{Filling the interior of the throats}
As seen in the previous section, we can solve elliptic equations with 
various boundary conditions in all the space exterior to two non-intersecting
 spheres $S_1$ and $S_2$. But a problem arises from the iterative nature of the total 
numerical procedure. Suppose that after a particular step the lapse 
$N=N_1 + N_2$ has been calculated by means of the two Poisson equations 
(\ref{e:split_lapse}). From the very procedure of the elliptic 
solvers, $N_1$ (resp. $N_2$) is known everywhere outside sphere $S_1$ 
(resp. $S_2$). If the next equation 
to be solved is the one for the shift vector split like 
(\ref{e:split_beta}), $N$ appears in the source term. We need to 
know the source everywhere outside 
the associated sphere $S_a$ ($a=1,2$) which includes the interior of the 
other sphere.
So we must construct fields that are known in the all space. After each 
resolution, the fields are filled as smoothly as possible inside the 
associated sphere. In our example, after the resolution of 
(\ref{e:split_lapse}), $N_1$ and $N_2$ are
filled inside the spheres, so that the total function $N$ is known everywhere.

The filling is performed, for each spherical harmonic $\left(l,m\right)$, by 
the following radial function~:
\begin{itemize}
\item $\left(3r^4-2r^6\right)\left(\alpha + \beta r^2\right)$ if $l$ is even,
\item  $\left(3r^4-2r^6\right)\left(\alpha r + \beta r^3\right)$ if $l$ is
odd,
\end{itemize}
where the coefficients $\alpha$ and $\beta$ are calculated so that the 
function is ${\mathcal C}^1$ across the sphere $S_a$. The 
multiplication by the polynomial $\left(3r^4-2r^6\right)$ 
ensures that the function is rather regular at the origin.
Of course this choice 
of filling is not unique and the final result should be independent of 
the filling procedure, the fields outside the spheres depending only on 
the boundary conditions on those spheres. The choice of filling may 
only change the convergence of 
the numerical scheme. Let us stress that even 
if the fields are known, regular 
and ${\mathcal C}^1$ everywhere, they have a physical meaning only outside 
the throats. The filling is only introduced for numerical purposes.

\subsection{Treatment of the extrinsic curvature tensor}

\subsubsection{Regularization of the shift}	\label{s:regul}

When one imposes corotation for the two black holes, that is a vanishing shift
vector on the throats, isometry conditions (59), 
(60) and (61) of Paper I are trivially
fulfilled. Unfortunately this is not the case for
(62) and (63) of Paper I. We must find a way to impose that
\be
\label{e:isometric}
\left. \frac{\partial \beta^\theta}{\partial r} \right|_{S_i} = 0
\quad {\rm and} \quad
\left. \frac{\partial \beta^\varphi}{\partial r} \right|_{S_i} = 0
\ee
in order to get a truly isometric solution.

Another problem comes from the computation of the reduced extrinsic curvature tensor
$\hat{A}^{ij}$ by means of 
(\ref{e:def_A}). Because of division by $N=0$ on $S_1$ and $S_2$, we must 
impose that
\be
\label{e:condition_regul}
\left. \left(L\beta\right)^{ij}\right|_{S_1} = 0 
{\rm \phantom{000}and\phantom{000}} 
\left. \left(L\beta\right)^{ij}\right|_{S_2} = 0,
\ee
so that the extrinsic curvature tensor is regular everywhere.
Because of the rigidity conditions (\ref{e:bound_beta}) and for a truly 
isometric solution verifying (\ref{e:isometric}), the regularity 
conditions (\ref{e:condition_regul}) are fulfilled if and only if
\be
\label{e:derivee_beta}
\left. \frac{\partial \beta^{r_1}}{\partial r_1}\right|_{S_1} = 0 
{\rm \phantom{000}and\phantom{000}}
\left. \frac{\partial \beta^{r_2}}{\partial r_2}\right|_{S_2} = 0.
\ee

So, to get a truly isometric and  regular solution, both the value and the 
radial derivative of $\vec{\beta}$ must be zero on the throats~:
\be
\left. \vec{\beta}\right|_{S_i} = 0
\quad {\rm and} \quad
\left. \frac{\partial \vec{\beta}}{\partial r}\right|_{S_i} = 0.
\ee

But when solving Eq. (\ref{e:eq_beta}), one can only impose the 
value at infinity and one of those two conditions, i.e. we can only solve 
for the Dirichlet or Neumann problem, not for both. We choose to solve the 
equation (\ref{e:eq_beta}) for the Dirichlet boundary condition : 
$\vec{\beta}=0$ on both spheres.
Doing so, the regularity conditions (\ref{e:derivee_beta}),
as well as the remaining isometry conditions (\ref{e:isometric}),
are not necessarily fulfilled. 
After each step we must modify the obtained shift vector to enforce
(\ref{e:derivee_beta}) and (\ref{e:isometric}). The part of 
the shift generated by the hole $1$ is modified by
\begin{eqnarray}
\label{e:regul}
\left. \beta^i_1 \right|_{\rm new} &=& \left. \beta^i_1 \right|_{\rm old} +
\beta^i_{{\rm cor}, 1}\\
\beta^i_{{\rm cor}, 1} &:=& -\frac{\left(R-r_1\right)^3\left(r_1-a_1\right)}
{\left(R-a_1\right)^3} \left.
\frac{\partial \left. \beta^i \right|_{\rm old}}{\partial r_1}
\right|_{S_1}   		\label{e:beta_cor_def}     \ ,
\end{eqnarray}
where $r_1$ is the radial coordinate associated with hole $1$, $a_1$ the 
radius of the throat and $R$ an arbitrary radius, typically 
$R = 2 a_1$. The correction procedure is only  applied for 
$a_1 \leq r_1 \leq R$. Let us mention that the function of $r_1$
in front of $\partial \left. \beta^i \right|_{\rm old} / \partial r_1$ 
in Eq.~(\ref{e:beta_cor_def}) has been chosen
so that it maintains the value of the shift vector on the sphere $1$ and its 
continuity (${\mathcal C}^1$ function). The same operation is done for the 
other hole. After regularization, the shift vector satisfies 
(i) the rigidity condition (\ref{e:bound_beta}), 
(ii) the isometry conditions (\ref{e:isometric}), and 
(iii) the condition (\ref{e:derivee_beta}) ensuring the regularity
of the extrinsic curvature, but it violates slightly the
momentum constraint~(\ref{e:eq_beta}).

As seen in Paper~I, the regularity is a consequence of the equation
\be
\label{e:div_beta}
\bar{D}_i \beta^i = -6\beta^i\bar{D}_i \ln \Psi.
\ee
Because this equation is not part of the system we choose to solve, we do not 
expect that the correction function is exactly zero at the end of a 
computation. But we will verify in Sec.~\ref{s:check_mom}
that it is only a small 
fraction of the shift vector (less than $10^{-3}$), fraction which 
represents the deviation from 
Eq. (\ref{e:div_beta}) (see also Ref.~\cite{Cook01} for an extended discussion). 
Moreover, we will see in Sec.~\ref{s:kerr}
that $\vec{\beta}_{\rm cor}$ converges to zero for a single rotating 
black hole.

\subsubsection{Computation of the extrinsic curvature tensor}

Using the regularized shift vector presented above, we 
can compute the tensor $\left(L\beta\right)^{ij}$, which is zero on both 
throats. To calculate the tensor $\hat{A}^{ij}$ one must divide it by the 
lapse function which also vanishes on both throats. Near the throat $1$, 
$N$ has the following behavior
\be
\left. N \right|_{r_1 \rightarrow a_1} = \left(r_1-a_1\right) n_1 \ ,
\ee
where $n_1$ is non zero on throat $1$ (this supposes that $r_1 = a_1$ is 
only a single pole of $N$, which turns out to be true, $\partial N / \partial r_1$ 
representing the ``surface gravity'' of black hole $1$). We can compute $n_1$,
using an operator that acts in the coefficient space of $N$ and divides it by
$ \left(r_1-a_1\right)$. The same operation is done with 
\be
\left. \left(L\beta\right)^{ij} \right|_{r_1 \rightarrow a_1} = 
\left(r_1-a_1\right) l_1^{ij}.
\ee
The divisions are also done on the second throat.
To compute the extrinsic curvature tensor in all space we use
\begin{itemize}
\item $\hat{A}^{ij} = l_1^{ij}/\left(2n_1\right)$ in the first shell
around throat $1$
\item $\hat{A}^{ij} = l_2^{ij}/\left(2n_2\right)$ in the first shell
around throat $2$
\item $\hat{A}^{ij} = \left(L\beta\right)^{ij}/\left(2N\right)$ in all other regions.
\end{itemize}
This procedure enables us to compute the extrinsic curvature tensor 
everywhere, without any problem that could arise from a division by zero.

\subsubsection{Splitting of the extrinsic curvature tensor}
In the split equations (\ref{e:split_lapse}) and (\ref{e:split_psi}), the 
term $\hat{A}_1^{ij}$ appears. This term represents the part of the total 
extrinsic curvature tensor generated mostly by hole $1$ so that the total 
tensor is given by
\be
\hat{A}^{ij}  = \hat{A}_1^{ij} + \hat{A}_2^{ij}.
\ee
For the binary neutron stars treated in \cite{GourgGTMB01}, those split quantities 
were constructed by setting $\hat{A}_1^{ij} = \left(L\beta_1\right)^{ij}/\left(2N\right)$.
Such a construction is not applicable in the case of black holes. 
Indeed, only the total shift vector is such that $\left(L\beta\right)^{ij}=0$ 
on the throats and not the split shifts $\vec{\beta}_1$ and $\vec{\beta}_2$. 
If such a construction were applied
the quantity $\hat{A}_1^{ij}$ would be divergent due to division by $N=0$ on
the throats. The computation presented in the previous section 
gets rid of such divergences but enables us to calculate only
the total $\hat{A}^{ij}$.

The construction of $\hat{A}_1^{ij}$ and $\hat{A}_2^{ij}$ is then obtained 
by 
\begin{eqnarray}
\hat{A}_1^{ij} &=& \hat{A}^{ij} H_1 \\
\hat{A}_2^{ij} &=& \hat{A}^{ij} H_2,
\end{eqnarray}
where $H_1$ and $H_2$ are smooth functions such that $H_1+H_2 = 1$ 
everywhere. We 
also want $H_1$ (resp. $H_2$) to be close to one near hole $1$ (resp. $2$) 
and close to zero near hole $2$ (resp. $1$), so that $\hat{A}_1^{ij}$ 
(resp. $\hat{A}_2^{ij}$) is mostly concentrated around hole $1$ (resp. $2$).
So, we define $H_1$ by
\begin{itemize}
\item $1$ if $r_1 \leq R_{\rm int}$
\item $\frac{1}{2}\left[\cos ^2\left(\pi/2
\left(r_1-R_{\rm int}\right)/
\left(R_{\rm ext}-R_{\rm int}\right)\right)+1\right]$ if 
$R_{\rm int}\leq r_1 \leq R_{\rm ext}$
\item $0$ if $r_2 \leq R_{\rm int}$ 
\item $\frac{1}{2}\sin^2\left(\pi/2
\left(r_1-R_{\rm int}\right)/
\left(R_{\rm ext}-R_{\rm int}\right)\right)$ if $R_{\rm int}\leq r_2 \leq R_{\rm ext}$
\item $\frac{1}{2}$ if $r_1 \ge R_{\rm ext}$ and $r_2 \ge R_{\rm ext}$,
\end{itemize}
where $r_1$ (resp. $r_2$) is the radial coordinate associated with throat
$1$ (resp. $2$). The radii $R_{\rm int}$ and $R_{\rm ext}$ are computational 
parameters, chosen so that the different cases presented above are exclusive.
Typically, we choose $R_{\rm int} = d/6$ and $R_{\rm ext} = d/2$, 
where $d$ is the coordinate distance between the centers of the throats.
$H_2$ is obtained by permutation of indices $1$ and $2$.

\subsection{Numerical implementation}

The numerical code implementing the method described above is written in
LORENE (Langage Objet pour la RElativit\'e Num\'eriquE),
which is a C++ based language for
numerical relativity developed by our group.
A typical run uses 12 domains (6 centered on each black hole)
and $N_r\times N_\theta \times
N_\varphi = 33\times 21 \times 20$ (resp.  $N_r\times N_\theta \times
N_\varphi = 21\times 17 \times 16$) coefficients in each domain in high 
resolution (resp. low resolution). For each value of 
$\Omega$, a typical calculation takes 50 steps. To determine the right 
value of the angular velocity, by means of a secant method, it takes usually 5 
different calculations with different values of $\Omega$. The associated time 
to calculate one configuration is approximatively 72 hours (resp. 36 hours)
for the high resolution (resp. for the low resolution) on one CPU of a 
SGI Origin200 computer (MIPS R10000 processor at 180 MHz). The corresponding 
memory requirement is 700~MB (resp. 300~MB) for the high resolution (resp. low
resolution).

\section{Tests passed by the numerical schemes}\label{s:tests}
\subsection{Schwarzschild black hole}
In this section we solve Eqs. (\ref{e:eq_lapse}) and (\ref{e:eq_psi}), with 
boundary conditions (\ref{e:bound_lapse}) and (\ref{e:bound_psi}) on
a single throat $S$. The behaviors at infinity are given by Eqs. 
(\ref{e:lapse_infinity}) and (\ref{e:psi_infinity}). In this particular 
case, the shift vector $\vec{\beta}$ is set to 
zero, so that $\hat{A}^{ij}$ vanishes. This represents a single, static black 
hole, and we expect to recover the Schwarzschild solution in isotropic 
coordinates.

The computation has been conducted with a initial guess far from the expected 
result. More precisely, we began the computation by setting $N=1$ and $\Psi=1$
everywhere. Equations  (\ref{e:eq_lapse}) and (\ref{e:eq_psi}) are 
then solved by iteration. Let us mention that the boundary condition
 on the conformal factor, given by (\ref{e:bound_psi}), is obtained by 
iteration. At each step we impose
\be
\left. \frac{\partial \Psi^J}{\partial r}\right|_S =
\left. \frac{\Psi^{J-1}}{2 r} \right|_S,
\ee
where $\Psi^J$ is the conformal factor at the current step and $\Psi^{J-1}$ 
at the previous one.

Before beginning a new step, some relaxation is performed on the fields by
\be
Q^J \leftarrow \lambda Q^J + (1-\lambda) Q^{J-1},
\ee
where $0 < \lambda \leq 1$ is the relaxation parameter, typically 
$\lambda = 0.5$. $Q$ stands for any of the fields for which we solve 
an equation ($N$ and $\Psi$ solely for the static case).

The iteration is stopped when the relative difference between the lapse 
obtained at 
two consecutive steps is smaller than the threshold $\delta N = 10^{-13}$. The 
computation has been performed with various number of collocation points and with 
two shells. All the errors are estimated by the infinite norm of the difference.

\begin{figure}
\centerline{\includegraphics[height=9cm,angle=-90]{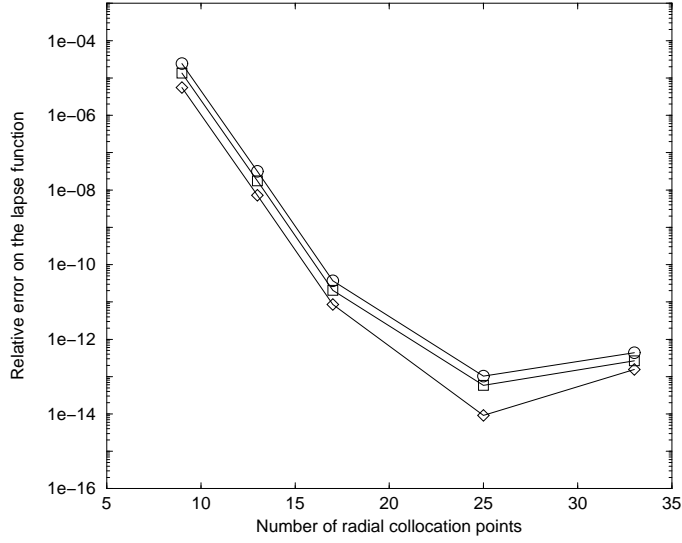}}
\vskip12pt
\caption{\label{f:erreur_n}
Relative difference between the calculated and the analytical lapse $N$ with 
respect to the number of radial spectral coefficients for the Schwarzschild 
black hole. The circles denote 
the error in the innermost shell, the squares that in the other shell and 
the diamonds that in the external domain.
}
\end{figure}

\begin{figure}
\centerline{\includegraphics[height=9cm,angle=-90]{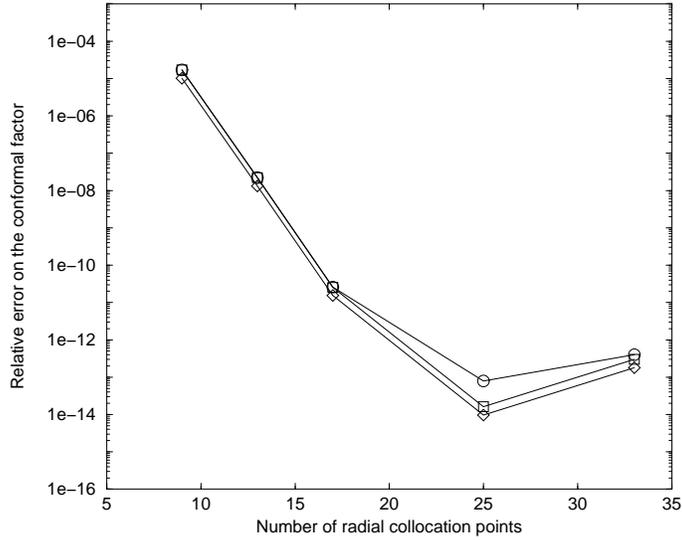}}
\vskip12pt
\caption{\label{f:erreur_psi}
Same as Fig. \ref{f:erreur_n} for the conformal factor $\Psi$.
}
\end{figure}

Figures \ref{f:erreur_n} and \ref{f:erreur_psi} show a extremely good agreement 
with the exact analytical solution. The saturation level of approximatively 
$10^{-13}$ is due to the finite number of digits (15) used in the 
calculations (round-off errors). Before the saturation, the error is evanescent 
(exponential decay with the number of collocation points), which is typical of 
spectral methods.

\subsection{Kerr black hole}\label{s:kerr}
In this section we consider a single rotating black hole by setting 
$\vec{\beta} \not= 0$. Let us mention that, since the Kerr solution is known to 
diverge from conformal flatness (see e.g. \cite{GaratP00}), 
we will no be able to recover exactly 
the Kerr metric. In other words the obtained solution is expected to violate 
some of the $5$ Einstein equations we decided to ignore.

So we solve Eqs. (\ref{e:eq_lapse}), (\ref{e:eq_beta}) and 
(\ref{e:eq_psi}) with boundary conditions  (\ref{e:bound_lapse}) 
(\ref{e:bound_beta}) and (\ref{e:bound_psi}) on one single sphere. The values at 
infinity are chosen to maintain asymptotical flatness by using Eq. 
(\ref{e:lapse_infinity}), (\ref{e:beta_infinity}) and (\ref{e:psi_infinity}).
The two parameters of our rotating black hole are the radius of the throat 
$S$ and the rotation velocity $\Omega$. The total mass $M$ and and angular 
momentum $J$ are computed at the end of the iteration.

Initialy the values of $N$ and $\Psi$ are set to those of a Schwarzschild 
black hole and the shift is set to zero. Relaxation is used for all the fields
with a parameter $\lambda = 0.5$. As for the Schwarzschild 
computation, we use two shells with the same number $N_r \times N_\theta \times N_\varphi$
of collocation points in the two shells and in the external compactified domain.
The iteration is stopped when the relative 
difference between the shifts obtained at two consecutive steps is smaller 
than $\delta \beta = 10^{-10}$.

\begin{figure}
\centerline{\includegraphics[height=9cm,angle=-90]{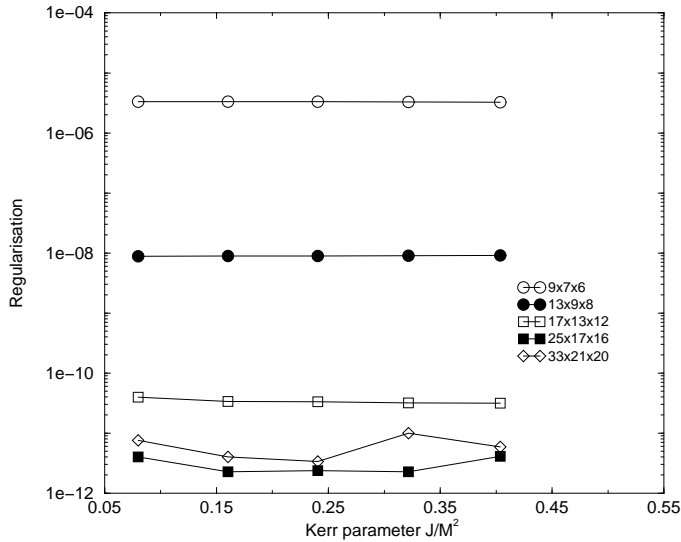}}
\vskip12pt
\caption{\label{f:regul_kerr}
Relative norm of the regularization function given by Eq. (\ref{e:regul}) with 
respect to the Kerr parameter $J/M^2$, for various numbers 
$N_r\times N_{\theta}\times N_{\varphi}$ of collocation points.
}
\end{figure}

Before comparing the obtained solution to the Kerr metric we perform some 
self-consistency checks, by varying the number of coefficients of the 
spectral expansion. First of all, we need to verify that the regularization
function applied to the shift by means of Eq. (\ref{e:regul}) has gone to zero at 
the end of the computation. Figure \ref{f:regul_kerr} shows that, for 
various values of the Kerr parameter $J/M^2$, 
the relative norm of the regularization function 
decreases very fast, as the number of coefficients increases. The saturation 
value of $10^{-11}$ is due to the criterium we choose to stop the computation 
$\delta \beta = 10^{-10}$. Had it been conducted for a greater number of 
steps, the saturation level of the double precision would have been reached.
Figure \ref{f:regul_kerr} enables us to say that the shift solution of 
(\ref{e:eq_beta}) fulfills the regularity conditions (\ref{e:bound_beta}) 
for the extrinsic curvature tensor. Let us mention, that the fact that the 
conformal approximation is not valid, does not prevent the correction function 
$\vec{\beta}_{\rm cor}$ from going to zero.

As seen in Paper I, the total angular momentum can be calculated in two 
different ways, using a surface integral at infinity:
\be
\label{e:J_inf_seul}
J = \frac{1}{8\pi}\oint_{\infty}\hat{A}^i_j m^j dS_i \ ,
\ee
(where $\w{m} := \partial / \partial \varphi$) or an integral on the
throat:
\be
\label{e:J_hor_seul}
J = -\frac{1}{8\pi} \oint_{r=a} \Psi^6 \hat{A}^{ij} f_{jk} m^k d\bar{S}_i   \ ,
\ee
where $ d\bar{S}_i$ denotes the surface element with respect to the flat 
metric $\w{f}$.

\begin{figure}
\centerline{\includegraphics[height=9cm,angle=-90]{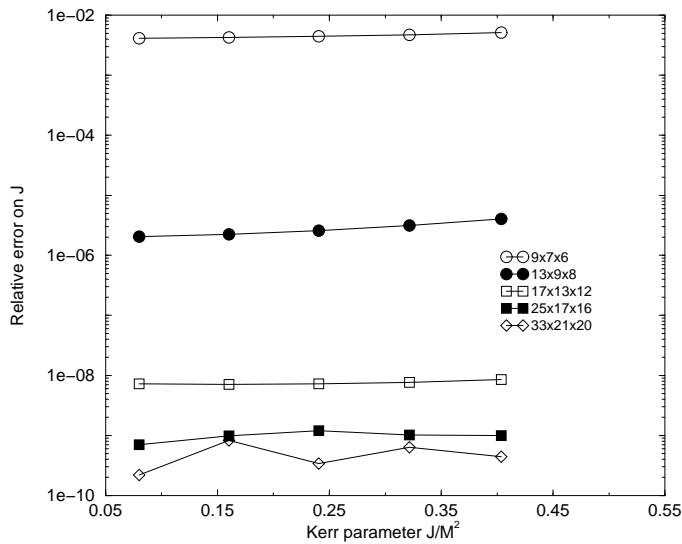}}
\vskip12pt
\caption{\label{f:test_j_kerr}
Same as Fig. \ref{f:regul_kerr} for the relative difference between the 
angular momentum calculated by means of Eq. (\ref{e:J_inf_seul}) and that by 
means of Eq. (\ref{e:J_hor_seul}).
}
\end{figure}

The two results will coincide if and only if the momentum constraint 
\be
\label{e:mom_contr}
\bar{D}_i\left(\Psi^6\hat{A}^{ij}\right) = 0
\ee
has been accurately solved in all the space. This is a rather strong test for 
the obtained value of $\hat{A}^{ij}$. Figure \ref{f:test_j_kerr} shows that
the relative difference between the two results
rapidly tends to zero, as the number of coefficients increases. The same 
saturation level as in Fig. \ref{f:regul_kerr} is observed.

\begin{figure}
\centerline{\includegraphics[height=9cm,angle=-90]{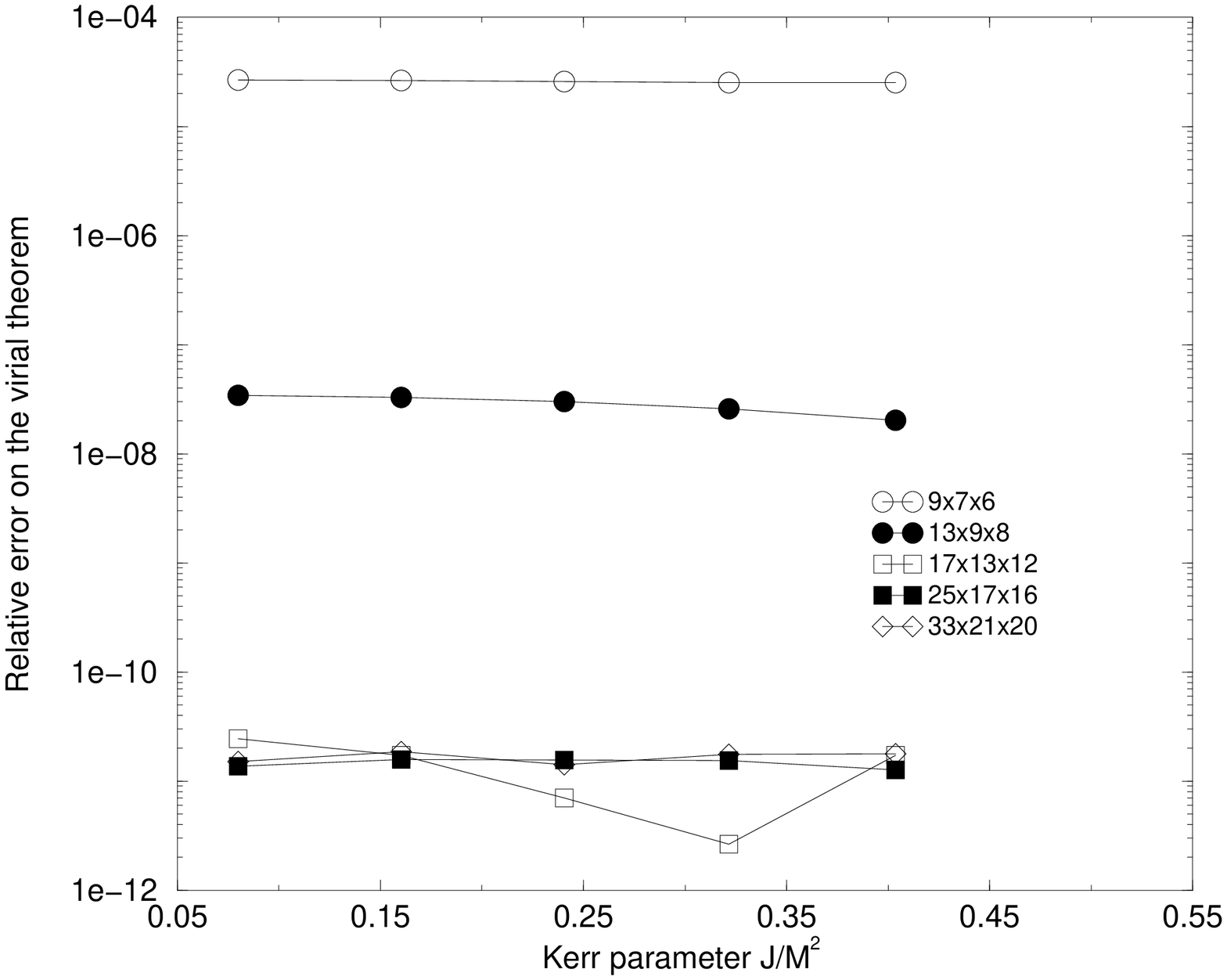}}
\vskip12pt
\caption{\label{f:viriel_kerr}
Same as Fig. \ref{f:regul_kerr} for the relative error on the virial theorem.
}
\end{figure}

The last self-consistency check is to verify the virial theorem 
considered in Sec. \ref{s:introduction}. In other words we wish to check 
if the ADM and Komar masses are identical, which should be the case for a Kerr
black hole. We plotted the relative difference between these two masses, for 
various numbers of collocation points and rotation velocities in Fig. 
\ref{f:viriel_kerr}. Once more this difference rapidly tends to zero as the 
number of coefficient increases. Contrary to the case of two black holes, 
the angular velocity $\Omega$ is not constrained by the virial theorem,
reflecting the fact that an isolated black hole can rotate at any velocity 
(smaller than the one of an extreme Kerr black hole).

To end with a single rotating black hole, we check how far the numerical
solution is from an exact, analytically given, Kerr black hole. Given the ADM 
mass $M$ and the reduced angular  momentum $a = J/M$ , an exact 
Kerr metric in quasi-isotropic coordinates would take the form
\be
ds^2 = -N_{\rm Kerr}^2 dt^2 + B_{\rm Kerr}^2 r^2 \sin^2 \theta 
\left(d\varphi-N_{\rm Kerr}
^\varphi dt\right)^2
	+ A_{\rm Kerr}^2 \left(dr^2+r^2d\theta^2\right),
\ee
with $N_{\rm Kerr}$, $N^\varphi_{\rm Kerr}$, $A_{\rm Kerr}$ and 
$B_{\rm Kerr}$ known functions. It obviously differs from 
asymptotical flatness because $A \not= B$ for $a \not= 0$. So we define a 
pseudo-Kerr metric 
by setting $B=A$, which gives
\be
ds^2 = -N_{\rm Kerr}^2 dt^2 + \Psi_{\rm Kerr}^4 \left[r^2 \sin^2 \theta 
\left(d\varphi-N_{\rm Kerr}^\varphi dt\right)^2
	+ \left(dr^2+r^2d\theta^2\right)\right],
\ee
where $\Psi^4_{\rm Kerr} = A^2_{\rm Kerr}$.
After a numerical calculation, we compute the global parameters $M$ and $a$,
calculate the functions $N_{\rm Kerr}$, $N^{\varphi}_{\rm Kerr}$ and 
$\Psi_{\rm Kerr}$ and 
compare them to the ones that have been calculated numerically. Note that 
$N^{\varphi}:=\beta^{\varphi}-\Omega$. The coefficients
of the pseudo-Kerr metric are given by
\begin{eqnarray}
\label{e:lapse_kerr}
N^2_{\rm Kerr} &:=& 1-\frac{2 M R}{\Sigma} +
	\frac{4a^2M^2R^2\sin^2\theta}{\Sigma^2\left(R^2+a^2\right)+2a^2
\Sigma M R \sin^2\theta} \\
\label{e:psi_kerr}
\Psi^4_{\rm Kerr} &:=& 1+\frac{2M}{r}+\frac{3M^2+a^2\cos^2\theta}{2r^2}
+ \frac{\left(M^2-a^2\right)M}{2r^3} + \frac{\left(M^2-a^2\right)^2}{16r^4} \\
\label{e:beta_kerr}
N^{\varphi}_{\rm Kerr} &:=& \frac{2aMR}{\Sigma\left(R^2+a^2\right)+2a^2MR\sin^2\theta},
\end{eqnarray}
where
\begin{eqnarray}
R &:=& r+\frac{M^2-a^2}{4r}+M \\
\Sigma &:=& R^2 + a^2 \cos^2\theta.
\end{eqnarray}

\begin{figure}
\centerline{\includegraphics[height=9cm,angle=-90]{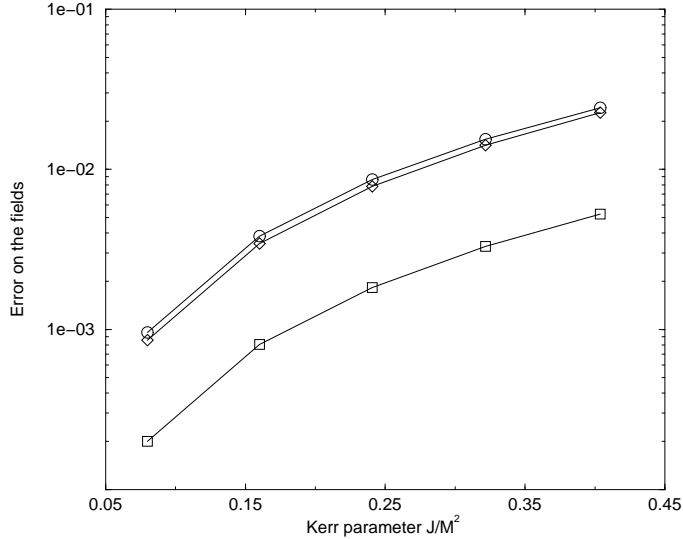}}
\vskip12pt
\caption{\label{f:compare_kerr}
Relative difference between the pseudo-Kerr quantities defined by
Eqs. (\ref{e:lapse_kerr})-(\ref{e:beta_kerr}) and the numerically 
calculated ones with respect to the angular velocity.
The computation has been performed with
$N_r \times N_{\theta} \times N_{\varphi} = 25 \times 17 \times 16$.
The circles denote the differences on $N$, the squares 
on $\Psi$ and the diamonds on $N^{\varphi}:=\beta^{\varphi}-\Omega$. 
}
\end{figure}

Those analytical functions are then compared with that obtained numerically 
(see Fig. \ref{f:compare_kerr}).
As expected the difference between the fields is not zero and it
increases with $\Omega$, reflecting the fact that a Kerr black hole deviates
more and more from conformal flatness as $J$ increases.

To summarize the results about a single rotating throat, we are confident in 
the fact that the Eqs. (\ref{e:eq_lapse}), (\ref{e:eq_beta}) and 
(\ref{e:eq_psi}) have been successfully and accurately solved, with 
the appropriate boundary conditions. On the other hand we do not claim to 
recover the exact Kerr metric, for this latter differs from conformal
flatness.

\subsection {The Misner-Lindquist solution}\label{s:misner}
Misner \cite{Misne63} and Lindquist \cite{Lindq63} have found the 
conformal factor $\Psi$ of two black holes in the static case, i.e. when 
$\vec{\beta}=0$ (see also Ref. \cite{Giuli98} and Appendices A and B of 
Ref. \cite{AndraP97}). In such a case the equation for $\Psi$ is only
\be
\label{e:psi_misner}
\Delta \Psi = 0,
\ee
which was to be solved using boundary conditions (\ref{e:psi_infinity}) 
and (\ref{e:bound_psi}). In the case of identical black holes, that is for 
two throats having the same radius $a$, the solution is analytical and does 
only depend on the separation parameter
\be
\label{e:def_separation}
D := \frac{d}{a} \ ,
\ee
$d$ being the coordinate 
distance between the centers of the throats. To check if our scheme enables us
to recover such a solution, we solve Eq. (\ref{e:psi_misner}) with the 
boundary conditions (\ref{e:psi_infinity}) and (\ref{e:bound_psi}). We then 
compute the ADM mass by means of the formula (see Paper I)
\be
M = -\frac{1}{2\pi} \oint_{\infty}\bar{D}^i\Psi dS_i
\ee
and compare the result to the analytical value given by a series in 
Lindquist article \cite{Lindq63}.

\begin{figure}
\centerline{\includegraphics[height=9cm,angle=-90]{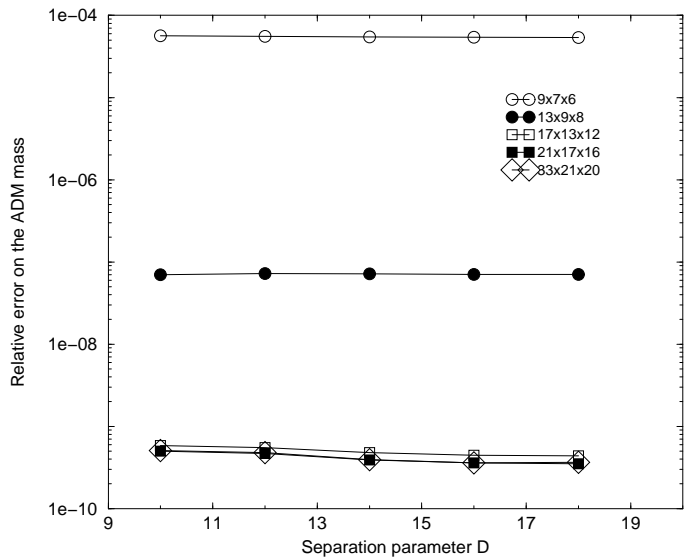}}
\vskip12pt
\caption{\label{f:ADM_misner}
Relative difference between the calculated and analytical ADM mass for 
the Misner-Lindquist solution.
The computation has been performed with various number of coefficients
$N_r \times N_{\theta} \times N_{\varphi}$.
}
\end{figure}

Let us mention that, even if Eq. (\ref{e:psi_misner}) is a linear equation
(the source is zero), the problem has to be solved by iteration because of 
our method for setting the boundary condition (\ref{e:bound_psi}).
The computation has been conducted with a relaxation parameter
$\lambda = 0.5$ and until a convergence of $\delta \Psi = 10^{-10}$ has been 
attained. The comparison between the analytical and calculated ADM
masses is plotted on Fig. \ref{f:ADM_misner} for various values of the 
separation parameter $D$ and various numbers of coefficients. The agreement is very
good for every value of $D$. As for the Kerr black hole, when the number of
coefficients increases, we attain the saturation level of a few $10^{-10}$ is 
due to 
the threshold chosen for stopping the calculation. This test makes us confident 
about the iterative scheme used to impose boundary conditions onto the two 
throats $S_1$ and $S_2$.

To go a a bit further and check the decomposition of the sources into two 
parts, presented in Sec. \ref{s:split}, we wish to consider a test problem 
with a source different from zero. To do so we consider the Misner-Lindquist
problem but decide to solve for the logarithm of $\Psi$, $\Phi = \ln \Psi$. 
The equation for $\Phi$ is 
\be
\label{e:equation_phi}
\Delta \Phi = -\bar{D}_k \Phi \bar{D}^k \Phi
\ee
and it must be solved with the following boundary conditions
\begin{eqnarray}
\Phi \rightarrow 0 &{\mathrm \phantom{000} when \phantom{000}}& r 
\rightarrow \infty \\
\left. \frac{\partial \Phi}{\partial r_1} \right|_{S_1} = -\frac{1}{2a_1} 
 &{\mathrm \phantom{000} and \phantom{000}}&
\left. \frac{\partial \Phi}{\partial r_2} \right|_{S_2} = -\frac{1}{2a_2}.
\end{eqnarray}

The source of the equation for $\Phi$ containing $\Phi$ itself, it is split 
as described in Sec. \ref{s:split} by
\be
\label{e:split_phi}
\Delta \Phi_a = -\bar{D}_k \Phi \bar{D}^k \Phi_a,
\ee
$a$ being $1$ or $2$. At the end of a computation, we compute the ADM mass 
by using
\be
M = -\frac{1}{2\pi} \oint_{\infty} \bar{D}^i \Phi dS_i
\ee
and compare it with the analytical value. The computation used a 
relaxation parameter $\lambda = 0.5$ and has been stopped 
when the threshold $\delta \Phi = 10^{-7}$ has been reached.

\begin{figure}
\centerline{\includegraphics[height=9cm,angle=-90]{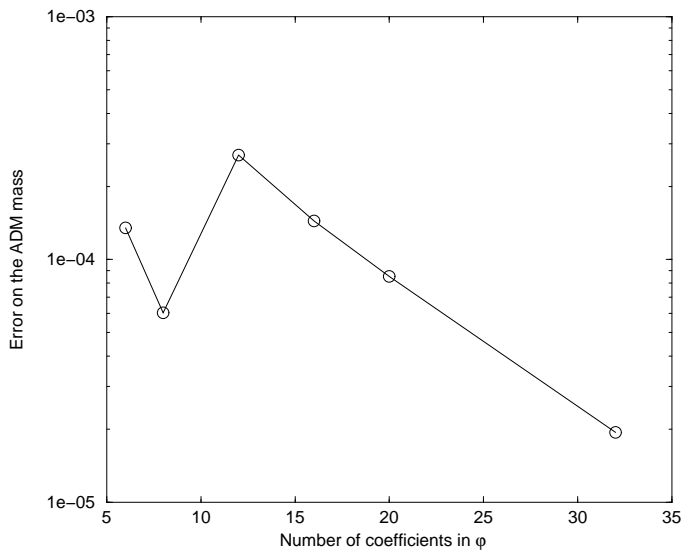}}
\vskip12pt
\caption{\label{f:ADM_phi}
Relative difference between the calculated and analytical ADM mass for 
the Misner-Lindquist solution calculated using $\Phi = \ln \Psi$, with 
respect to the number of coefficients in $\varphi$ ($N_\theta = N_\varphi+1$
and $N_r = 2N_\varphi+1$). The separation parameter is $D=10$.
}
\end{figure}

Figure \ref{f:ADM_phi} shows the resulting relative error
estimated by means of the ADM 
mass for $D=10$ and various numbers of coefficients. The convergence is 
{\em evanescent}, i.e. it is exponential as the number of coefficients increases.
Unfortunately, this convergence is much slower than when the solution was computed
using $\Psi$ and Eq. (\ref{e:psi_misner}). This comes from the very nature of the 
source of Eq. (\ref{e:split_phi}). The part of the equation split on 
the coordinates centered around throat $1$ is the sum of two terms.
The first one $-\bar{D}_k \Phi_1 \bar{D}^k \Phi_1$ is centered around hole $1$ and well 
described by spherical coordinates associated with this hole. We do not expect 
any problems with this term. The other term is 
$-\bar{D}_k \Phi_1 \bar{D}^k \Phi_2$ and contains a part that is centered around 
hole $2$. Describing this part using spherical coordinates around hole $1$ is much
more tricky and a great number of coefficients, especially in $\varphi$,
is necessary to do it accurately. It is the presence of such a component at the 
location of the other hole that makes the convergence of the calculation much 
slower in this case. Of course, we expect to recover this effect in the 
calculation of orbiting black holes.

\section{Sequence of equal mass corotating black holes in circular orbit}
\label{s:sequence}
\subsection{Numerical procedure}
\begin{figure}
\centerline{\includegraphics[height=9cm,angle=-90]{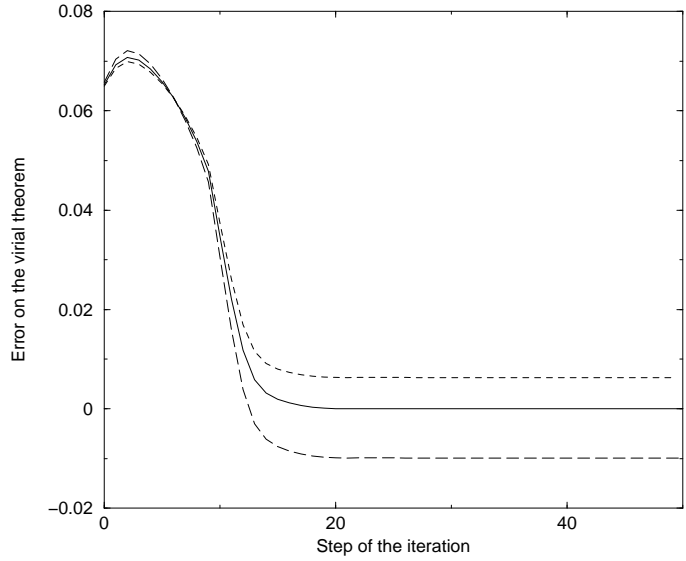}}
\vskip12pt
\caption{\label{f:find_omega}
Value of $\left(M_{\rm ADM}-M_{\rm Komar}\right)/M_{\rm Komar}$ with respect 
to the step of 
the iteration, for $D=16$ and for various values of $\Omega$. The solid line denotes
$\Omega_{\rm true}$, the short-dashed line $\Omega = 0.95\, \Omega_{\rm true}$ and the 
long-dashed line $\Omega = 1.08\, \Omega_{\rm true}$.
}
\end{figure}

In this section we concentrate on equal mass black holes. The only parameter 
is the ratio $D$ between the distance of the centers of the holes and the 
radius of the throats [see Eq.~(\ref{e:def_separation})].
We solve Eqs. (\ref{e:eq_lapse}), (\ref{e:eq_beta})
and (\ref{e:eq_psi}), with values at infinity given by (\ref{e:lapse_infinity}),
(\ref{e:beta_infinity}) and (\ref{e:psi_infinity}) and with boundary conditions 
on the horizons by (\ref{e:bound_lapse}), (\ref{e:bound_beta}) and 
(\ref{e:bound_psi}). We solve for various values of $\Omega$ and choose for 
solution the only value that fulfills the condition (\ref{e:condition_omega}). It
turns out that this process uniquely determines the angular velocity. Let us call 
$\Omega_{\rm true}$ the only value that equals the ADM and the Komar-like masses.
It happens that
\begin{itemize}
\item if $\Omega < \Omega_{\rm true}$ then $M_{\rm Komar} < M_{\rm ADM}$
\item if $\Omega > \Omega_{\rm true}$ then $M_{\rm Komar} > M_{\rm ADM}$.
\end{itemize}
The fact that $\Omega - \Omega_{\rm true}$ has always the same sign than
$M_{\rm Komar}-M_{\rm ADM}$ enables us to implement a 
very efficient procedure to determine 
the orbital velocity. It is found as the zero of the function 
$M_{\rm Komar}\left(\Omega\right)-M_{\rm ADM}\left(\Omega\right)$ by means of a secant 
method. This is illustrated by Fig. \ref{f:find_omega}, which shows the 
value $\left(M_{\rm ADM}-M_{\rm Komar}\right)/M_{\rm Komar}$ for various 
values of $\Omega$, 
with respect to the step of the iterative procedure. Those calculations have been
performed for $D=16$. The solid line denotes $\Omega_{\rm true}$, the only value 
of $\Omega$ for which $\left(M_{\rm ADM}-M_{\rm Komar}\right)/M_{\rm Komar}$ 
converges to $0$.

The computations have been done either in {\em low resolution} with 
$N_r \times N_\theta \times N_\varphi = 21\times 17 \times 16$ coefficients 
in each domain or in {\em high resolution} with 
$N_r \times N_\theta \times N_\varphi = 33\times 21 \times 20$ coefficients in each
domain.
All the computation used a relaxation parameter $\lambda = 0.5$. We solve first 
for the static case $\Omega = 0$ and use that solution as initial guess. For 
each value of $\Omega$, the computation is stopped for a relative change on the 
shift vector as small as $\delta \beta = 10^{-8}$ (resp. $\delta \beta = 10^{-7}$) 
for the high resolution (resp. low resolution) between two consecutive steps.
The secant procedure for the determination of the angular velocity has been 
conducted until $\left|\left(M_{\rm ADM}-M_{\rm Komar}\right)/M_{\rm Komar}\right| 
< 10^{-5}$ (resp. $< 10^{-4}$) for the high resolution (resp. low resolution), 
which gives a precision on $\Omega_{\rm true}$ of the order of $10^{-4}$ 
(resp. $10^{-3}$).

\subsection{Tests}  \label{s:tests_sequence}

\subsubsection{Check of the momentum constraint} \label{s:check_mom}

As discussed in Sec.~\ref{s:regul}, we have to slightly 
modify the shift vector to ensure both the regularity of the extrinsic
curvature on the throats
and the invariance of the shift under the inversion isometry. 
This modification of the shift, via the addition of the correction
function $\vec{\beta}_{\rm cor}$, results in a slight
violation of the momentum constraint Eq.~(\ref{e:eq_beta}). 
A good way to measure the magnitude of this
violation is to check whether the total angular momentum $J$ has the same value 
when calculated by surface integrals at infinity or on the throats. 
Indeed, as for the Kerr black hole, it has been 
shown in Paper I that $J$ can be given either by one of the two
following integrals:
\begin{eqnarray}
\label{e:J_inf}
J &=& \frac{1}{8\pi} \oint_{\infty}\hat{A}^i_j m^j d\bar{S}_i \ , \\
\label{e:J_hor}
J &=& -\frac{1}{8\pi}\oint_{S_1} \Psi^6 \hat{A}^{ij}f_{jk}m^k d\bar{S}_i
-\frac{1}{8\pi}\oint_{S_2} \Psi^6 \hat{A}^{ij}f_{jk}m^k d\bar{S}_i.
\end{eqnarray}
Any difference between those two formulas would reflect the fact that 
the momentum constraint (\ref{e:mom_contr}) is not exactly fulfilled.

\begin{figure}
\centerline{\includegraphics[height=8cm]{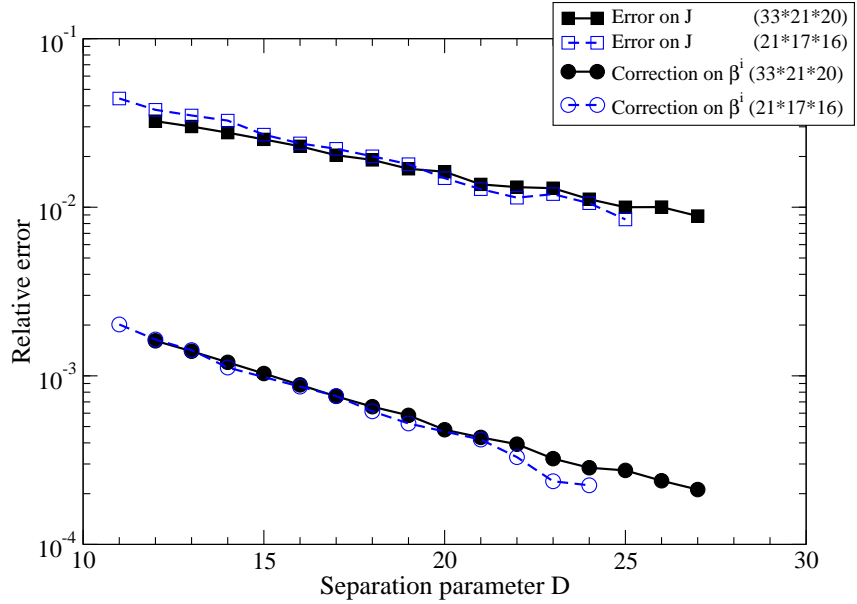}}
\vskip12pt
\caption{\label{f:verif_J}
Relative difference between the regularized shift and the
exact solution of (\ref{e:eq_beta}) (circles) and relative difference
between $J$ calculated by means of (\ref{e:J_inf}) and (\ref{e:J_hor})
(squares). 
The filled symbols and the solid line denote the high resolution and 
the empty symbols and the dashed line the low resolution.
}
\end{figure}

We have plotted the relative difference $\delta J / J$
between the two integrals 
(\ref{e:J_inf}) and (\ref{e:J_hor}) in Fig.~\ref{f:verif_J} as a function
on the separation between the two holes. Also shown on the same figure
is the relative norm of the correction function
$|\vec{\beta}_{\rm cor}|/|\vec{\beta}|$. 
The correlation between the two curves shows 
that the error on the momentum constraint arises from the 
introduction of the correction function on the shift. 
It is also clear from Fig.~\ref{f:verif_J} that increasing the number of
coefficients in the spectral method does not make
the correction function tend to zero. This means that the error in 
the momentum constraint come rather from the method 
(necessity to regularize the shift vector) than from some lack of 
numerical precision.

As discussed in Paper~I, we had to regularize the shift vector
because Eq.~(\ref{e:div_beta}) is not enforced in our scheme. 
It has been argued recently by Cook~\cite{Cook01} that if one reformulates
the problem by assuming that the helical vector $\wg{\ell}$ is not an
exact Killing vector, but only an approximate one --- as it is in reality ---
then the only freely specifiable part of the extrinsic curvature, 
as initial data, is (\ref{e:def_A}), not (\ref{e:div_beta}). This means that
the relation (\ref{e:div_beta}) between the extrinsic curvature and
the shift is not as robust as the relation (\ref{e:def_A}).

However, we see from Fig.~\ref{f:verif_J} that
at the innermost stable circular orbit, which is located at $D=17$ 
(cf. Sec.~\ref{s:evol_seq}), the error are very small:
\begin{eqnarray}
	\delta J / J & = & 2\ 10^{-2} \\
	|\vec{\beta}_{\rm cor}| & = & 8\ 10^{-4} |\vec{\beta}| \ . 
\end{eqnarray}
The $\delta J / J$ error estimator maximizes the error on the momentum
constraint because it integrates it in all space. 
Thus we conclude that momentum constraint is satisfied in our numerical
results with a precision of the order $1\%$.

\subsubsection{Check of the Smarr formula}

\begin{figure}
\centerline{\includegraphics[height=8cm]{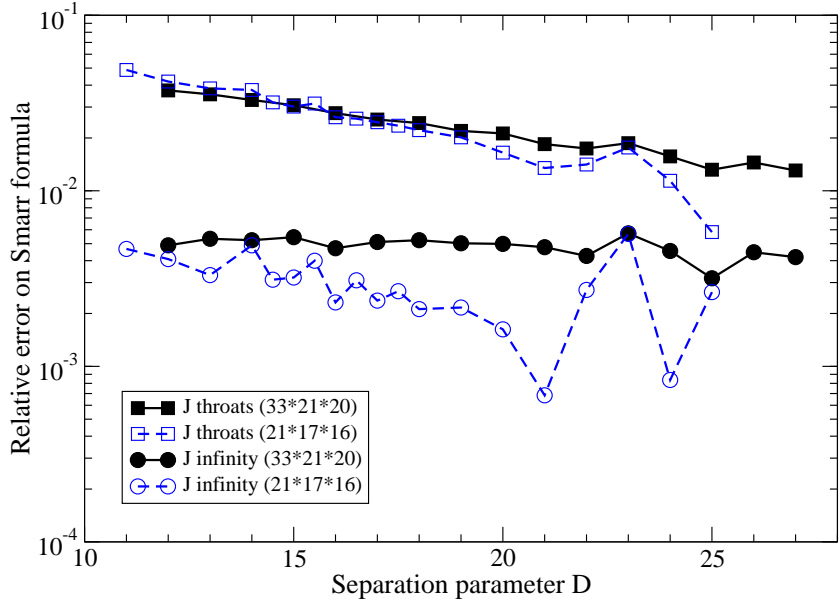}}
\vskip12pt
\caption{\label{f:smarr}
Relative error on the generalized Smarr formula (\ref{e:smarr}).
The circles denote the error obtained 
using $J$ calculated at infinity [Eq. (\ref{e:J_inf})] and the squares 
that obtained when evaluating $J$ on the throats [Eq. (\ref{e:J_hor})].
The filled symbols and the solid line denote the high resolution and the empty symbols
and the dashed line the low resolution.
}
\end{figure}

A good check of the global error in the numerical solution is
the generalized Smarr formula derived in Paper I~: 
\be
\label{e:smarr}
M-2\Omega J = -\frac{1}{4\pi}\oint_{S_1} \Psi^2 \bar{D}_i N d\bar{S}^i
	-\frac{1}{4\pi}\oint_{S_2} \Psi^2 \bar{D}_i N d\bar{S}^i.
\ee
For any computation, one gets $M$, $\Omega$ and can compute the r.h.s. of 
Eq. (\ref{e:smarr}) and use that equation to derive the value of $J$ that 
fulfills the Smarr formula. That value is then compared to the ones calculated 
using Eqs. (\ref{e:J_inf}) and (\ref{e:J_hor}). The comparison is plotted in 
Fig. \ref{f:smarr} for the two different resolutions.
It turns out that the angular momentum 
calculated at infinity is better in fulfilling the Smarr formula
than the one calculated on the throats by an 
order of magnitude and that the precision is better than $5 \times 10^{-3}$. So, 
for all following purposes, we will use the value of $J$ given by Eq. 
(\ref{e:J_inf}).

\subsubsection{Check of Kepler law at large separation}

The next thing one wishes to test is the value of $\Omega$, obtained 
from the virial criterium (\ref{e:condition_omega}). 
In Newtonian gravity, two points particles on 
circular orbits obey the following relation, which is
equivalent to Kepler's third law:
\be
\frac{4 J \Omega^{1/3}} {M^{5/3}} = 1,
\ee
where $M$ is the total mass, $J$ the total angular momentum and $\Omega$ the
orbital velocity. For large separations of the two throats we expect to recover
this relation. Therefore, for every value of $D$, we evaluate 
\be
\label{e:def_I}
I := \frac{4 J \Omega^{1/3}} {M^{5/3}}
\ee 
and check if $I$ tends to $1$ when $D\rightarrow \infty$.

\begin{figure}
\centerline{\includegraphics[height=8cm]{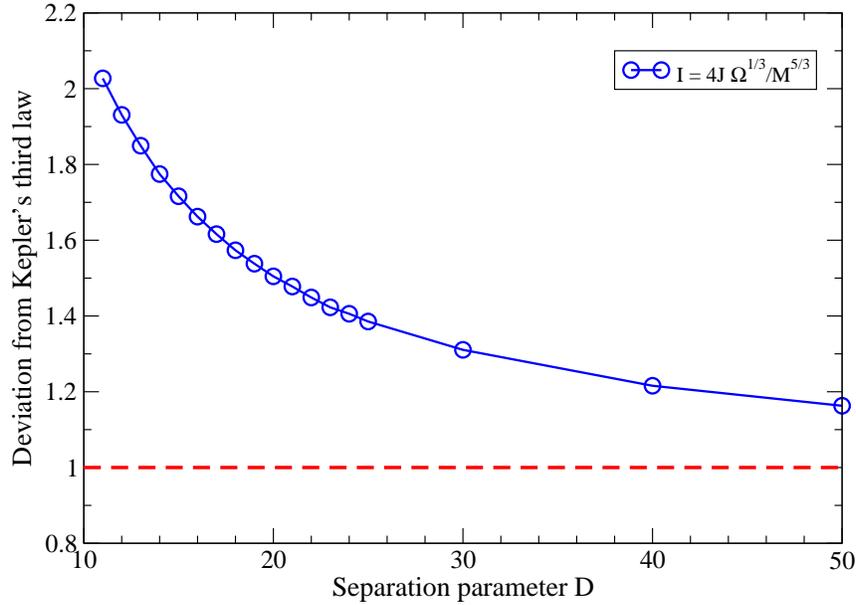}}
\vskip12pt
\caption{\label{f:kepler}
Value of $I=4J\left(\Omega/M^5\right)^{1/3}$ (low resolution runs)
with respect to the separation parameter $D$. 
The horizontal dashed line corresponds to the value predicted by
Kepler's third law. 
}
\end{figure}

The value of $I$ is plotted in Fig. \ref{f:kepler} with respect to the distance 
parameter $D$. As expected, for large values of $D$, it tends to $1$, implying 
that for large separations the system behaves like two point particles in Keplerian 
motion.

\subsection{Evolutionary sequence} \label{s:evol_seq}

\begin{figure}
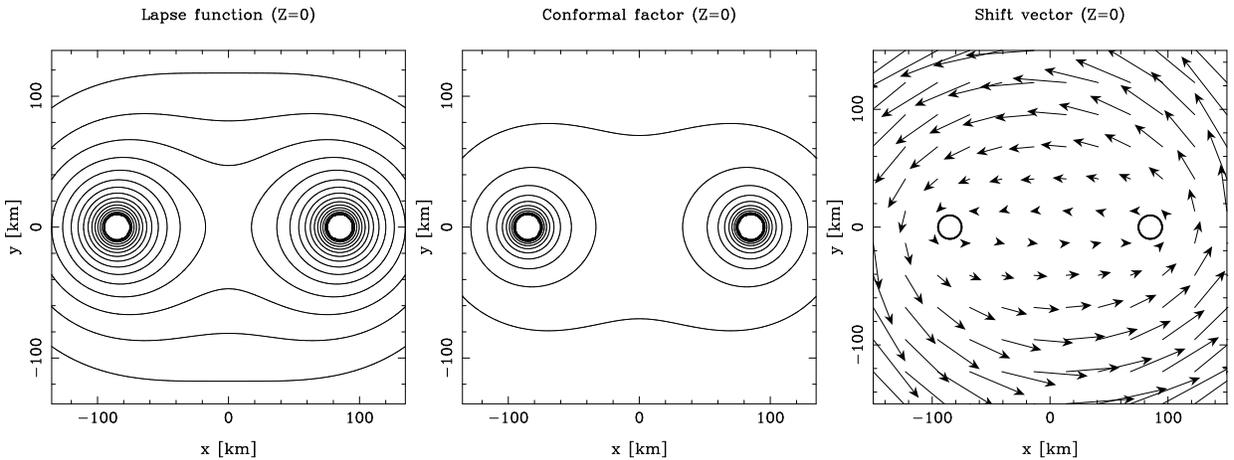

\centerline{\includegraphics[height=6cm]{lapse.eps}
	    \includegraphics[height=6cm]{psi.eps}
	    \includegraphics[height=6cm]{shift.eps}}
\vskip12pt
\caption{\label{f:lapse_psi_z}
Isocontour of the lapse function $N$ and of the conformal factor $\Psi$ and 
plot of the
shift vector $\vec{\beta}$, for $D=17$,  
in the orbital plane $z=0$. The computation has been done using the high resolution.
The thick solid lines denote the surfaces of the throats. The kilometer scale
of the axis corresponds to an ADM mass of $31.8 M_\odot$.
}
\end{figure}

\begin{figure}
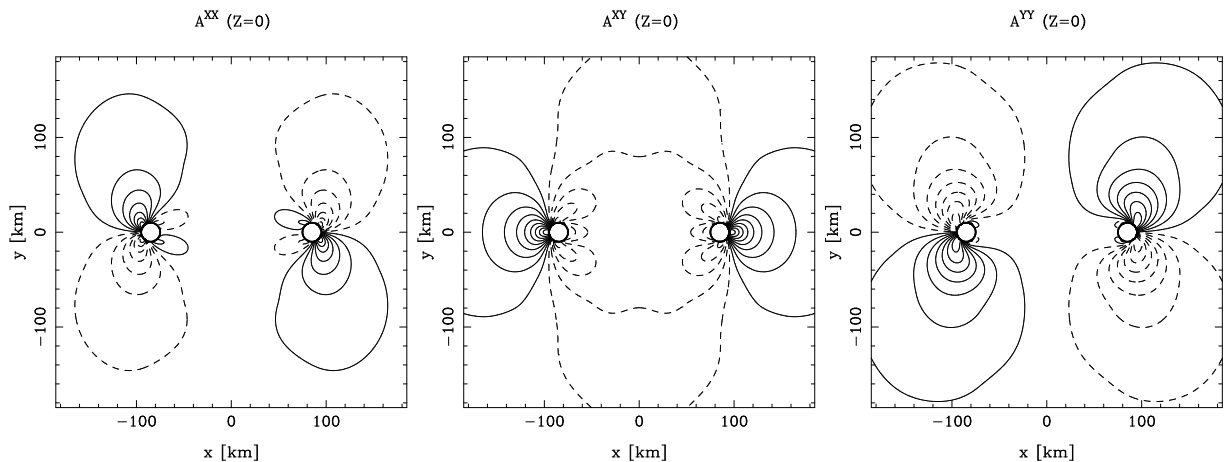

\centerline{\includegraphics[height=6cm]{kxx.eps}
	    \includegraphics[height=6cm]{kxy.eps}
	    \includegraphics[height=6cm]{kyy.eps}}
\vskip12pt
\caption{\label{f:k_z}
Isocontour of the extrinsic curvature tensor for $D=17$ in the orbital 
plane $z=0$.
The solid (dashed) lines denote positive (negative) values.
The thick solid lines denote the surfaces of the throats.
The computation has been done using the high resolution.
The kilometer scale
of the axis corresponds to an ADM mass of $31.8 M_\odot$.
}
\end{figure}

Let us first present some figures about the metric fields.
Figure \ref{f:lapse_psi_z} shows the total lapse function $N$, conformal factor
$\Psi$ and the shift vector $\vec{\beta}$ and Fig. \ref{f:k_z} the components 
$\hat{A}^{xx}$, $\hat{A}^{xy}$ and $\hat{A}^{yy}$
of the extrinsic curvature tensor. All those plots
are cross-section in the orbital plane $z=0$ and the coordinate system is
a Cartesian one centered at the middle of the centers of the throats. The computation
has been done using the high resolution. The separation 
parameter is $D=17$. As it will be seen later, this separation corresponds to the 
turning point in the energy and angular momentum curves.

In the previous section, the only parameter we considered was the dimensionless 
separation parameter $D$. But there also exists a scaling factor.
Suppose that all the distances 
in the computation are multiplied by some factor $\alpha$. Another solution with 
the same value of $D$ will be obtained, the global quantities being rescaled as
\begin{eqnarray}
M_{\alpha} &=& \alpha M_1 \\
J_{\alpha} &=& \alpha^2 J_1 \\
\Omega_{\alpha} &=& \frac{\Omega_1}{\alpha},
\end{eqnarray}
where $M_1$, $J_1$ and $\Omega_1$ are the values before rescaling and $M_{\alpha}$, 
$J_{\alpha}$ and $\Omega_{\alpha}$ the values after the rescaling.

Consider a physical configuration corresponding to a value 
$D\left(n\right)$ of the separation 
parameter, with global quantities $M\left(n\right)$, $J\left(n\right)$ and 
$\Omega\left(n\right)$. This system will evolve due to the emission of gravitational 
radiation. A subsequent configuration $n+1$ will have $D\left(n+1\right)<D\left(n\right)$.
But what scaling factor $\alpha$ should be applied to the
configuration calculated for $D\left(n+1\right)$ to ensure that it represents the
same physical system as before ? In other word, a physical sequence is a one parameter 
(the separation) family of configurations and we have to impose another condition
to determine the scaling factor associated with each value of $D$. 
In the case of binary neutron stars the condition is obtained by imposing that
the number of baryons is conserved (see e.g. Ref. \cite{GourgGTMB01}).
This cannot be extended to the black holes case since no
matter is present. We chose instead to define a sequence by requiring that
the loss of energy (ADM mass) $dM$ and angular momentum $dJ$
due to gravitational wave emission are related by
\be
\label{e:def_sequence}
\left.\frac{dM}{dJ}\right|_{\rm sequence} = \Omega.
\ee
This relation is exact
at least when one considers only the quadrupole formula (see e.g. page~478 of
Ref.~\cite{ShapiT83}). It turns out that it is
also well verified for sequences of binary neutron stars 
\cite{BaumgCSST98,UryuSE00}. So
Eq.~(\ref{e:def_sequence}) should hold rather well for corotating black holes.

The scaling factor $\alpha$ associated with the separation parameter 
$D\left(n+1\right)$ can be computed from the global values at separation 
$D\left(n\right)$ and the unscaled values at separation $D\left(n+1\right)$ 
as the solution the third degree equation
\be
\frac{M\left(n\right)-\alpha M_1\left(n+1\right)}
{J\left(n\right)-\alpha^2 J_1\left(n+1\right)} = 
\frac{1}{2}\left(\Omega\left(n\right)+
\frac{\Omega_1\left(n+1\right)}{\alpha}\right),
\ee
which is a first-order translation of Eq. (\ref{e:def_sequence}).
To present the results, we define the following dimensionless
quantities 
\begin{eqnarray}
\bar{M} := \frac{M}{M_0}\\
\bar{J} := \frac{J}{M_0^2}\\
\bar{\Omega} := M_0 \Omega\\
\bar{l} := \frac{l}{M_0},
\end{eqnarray}
where $l$ is the proper separation of the holes, defined as the geometrical distance 
between the throats along the axis joining their centers.
$M_0$ is some arbitrary mass used for normalization purpose. It is often
convenient to choose $M_0$ to be the total mass of the system when the two 
holes are infinitely separated, i.e. the ADM mass when $D \rightarrow \infty$. Unlike 
other methods, this value is not an input parameter of our calculation. It
can only be obtained by constructing a sequence until very large values 
of $D$, which would impose to calculate a great number of configurations.
However, as will been seen further, the system will exhibit turning point in the 
total energy and angular momentum, thereafter assumed to be the signature of an 
innermost stable circular orbit (thereafter ISCO). We chose $M_0$ to be the total ADM 
mass of the system at that point : 
\be
M_0 := \left.M_{\rm ADM}\right|_{\rm ISCO},
\ee
so that $\bar{M}$ is $1$ at the location of the ISCO.

The values of the dimensionless quantities $\bar{\Omega}$, $\bar{J}$, $\bar{M}$
and $\bar{l}$
along the sequence are given by Table~\ref{t:values}, for the high resolution.

\begin{table}
\caption{Values of dimensionless quantities along a sequence of corotating
black holes obtained using the high resolution. The bold line denotes the values at the 
location of the ISCO.}
 \begin{center}
  \begin{tabular}{ccccc}
  $D$ & $\bar{\Omega}$ & $\bar{J}$ & $\bar{M}$ & $\bar{l}$
  \\ \hline
$40$   &   $0.0296159$   &   $0.995862$   &   $1.00597$	&	$13.3502$\\
$39$   &   $0.0307026$   &   $0.987981$   &   $1.00573$	&	$13.0699$\\
$38$   &   $0.0318074$   &   $0.978701$   &   $1.00544$	&	$12.7892$\\
$37$   &   $0.0330632$   &   $0.971834$   &   $1.00522$	&	$12.5075$\\
$36$   &   $0.0344638$   &   $0.966535$   &   $1.00504$	&	$12.2247$\\
$35$   &   $0.0358923$   &   $0.959407$   &   $1.00479$	&	$11.9414$\\
$34$   &   $0.0374279$   &   $0.952494$   &   $1.00453$	&	$11.6572$\\
$33$   &   $0.0390612$   &   $0.945264$   &   $1.00426$	&	$11.3721$\\
$32$   &   $0.0408436$   &   $0.938754$   &   $1.004$	&	$11.086$\\
$31$   &   $0.0427491$   &   $0.931913$   &   $1.00371$	&	$10.7988$\\
$30$   &   $0.0448273$   &   $0.925596$   &   $1.00343$	&	$10.5104$\\
$29$   &   $0.0470335$   &   $0.918635$   &   $1.00311$	&	$10.2211$\\
$28$   &   $0.0494625$   &   $0.911941$   &   $1.00279$	&	$9.93012$\\
$27$   &   $0.0521747$   &   $0.907183$   &   $1.00255$	&	$9.6379$\\
$26$   &   $0.0550996$   &   $0.901696$   &   $1.00225$	&	$9.34411$\\
$25$   &   $0.0582842$   &   $0.896173$   &   $1.00194$	&	$9.04881$\\
$24$   &   $0.0617501$   &   $0.89047$   &   $1.0016$	&	$8.75188$\\
$23$   &   $0.0656222$   &   $0.885511$   &   $1.00128$	&	$8.45286$\\
$22$   &   $0.0699629$   &   $0.88151$   &   $1.00101$	&	$8.15164$\\
$21$   &   $0.0747426$   &   $0.877312$   &   $1.00071$	&	$7.84823$\\
$20$   &   $0.0801137$   &   $0.874079$   &   $1.00046$	&	$7.54243$\\
$19$   &   $0.086184$   &   $0.871511$   &   $1.00024$	&	$7.23364$\\
$18$   &   $0.0930453$   &   $0.869573$   &   $1.00007$	&	$6.9218$\\
${\bf 17}$   &   ${\bf 0.100897}$   &   ${\bf 0.86885}$   &   ${\bf 1}$	
&		${\bf 6.60644}$\\
$16$   &   $0.109958$   &   $0.869769$   &   $1.0001$	&	$6.28724$\\
$15$   &   $0.120329$   &   $0.870729$   &   $1.00021$	&	$5.96358$\\
$14$   &   $0.132657$   &   $0.874838$   &   $1.00073$	&	$5.63471$\\
$13$   &   $0.14712$   &   $0.880134$   &   $1.00147$	&	$5.30006$\\
$12$   &   $0.164512$   &   $0.888448$   &   $1.00276$	&	$4.95863$\\
\end{tabular}
 \end{center}
 \label{t:values}
\end{table}

\begin{figure}
\centerline{\includegraphics[height=9cm,angle=-90]{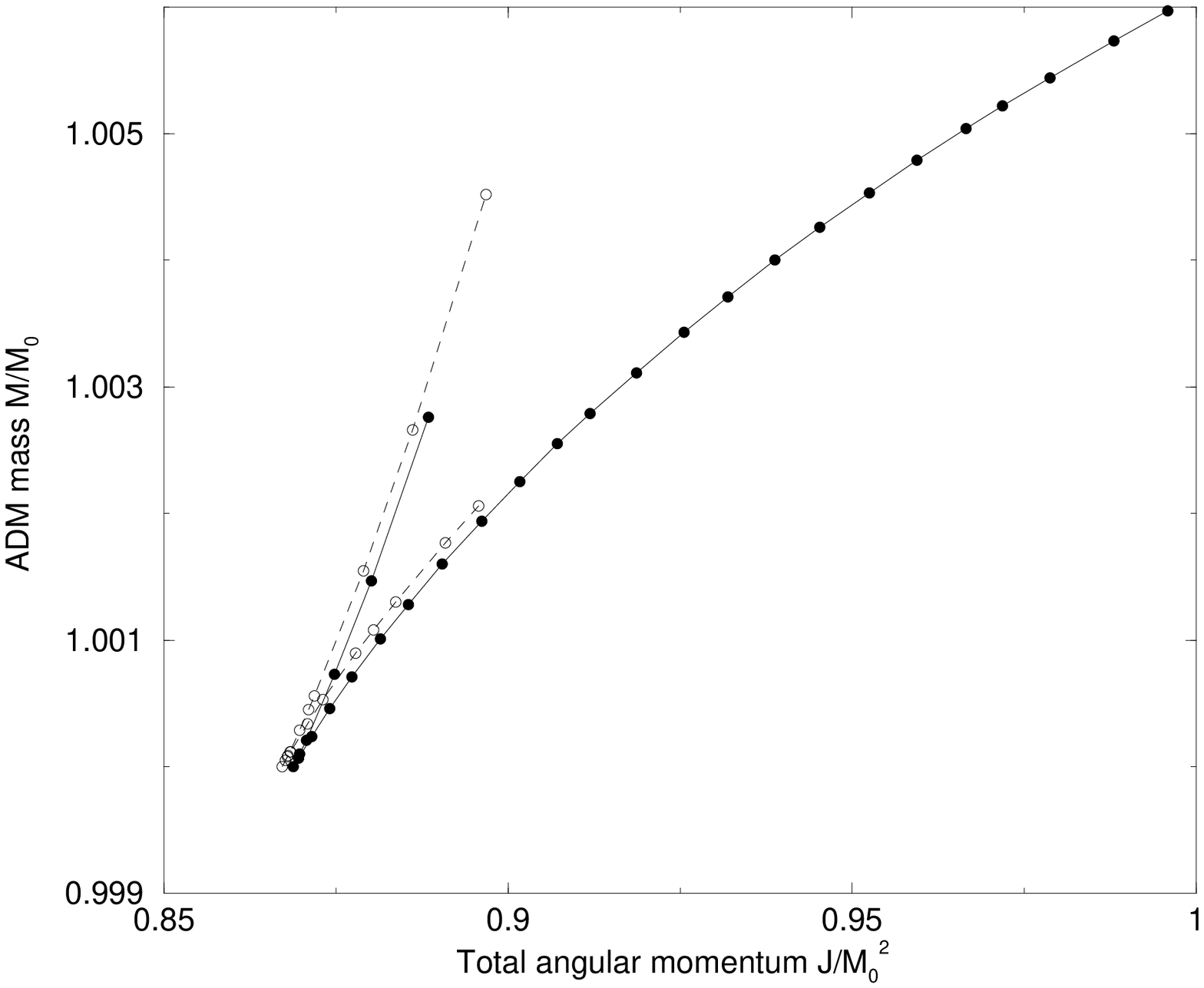}}
\vskip12pt
\caption{\label{f:e_f_j}
$\bar{M}$ with respect to $\bar{J}$ along a sequence. The filled symbols and solid
line denote the high resolution and the empty symbols and the dashed line the low one.
}
\end{figure}

\begin{figure}
\centerline{\includegraphics[height=9cm,angle=-90]{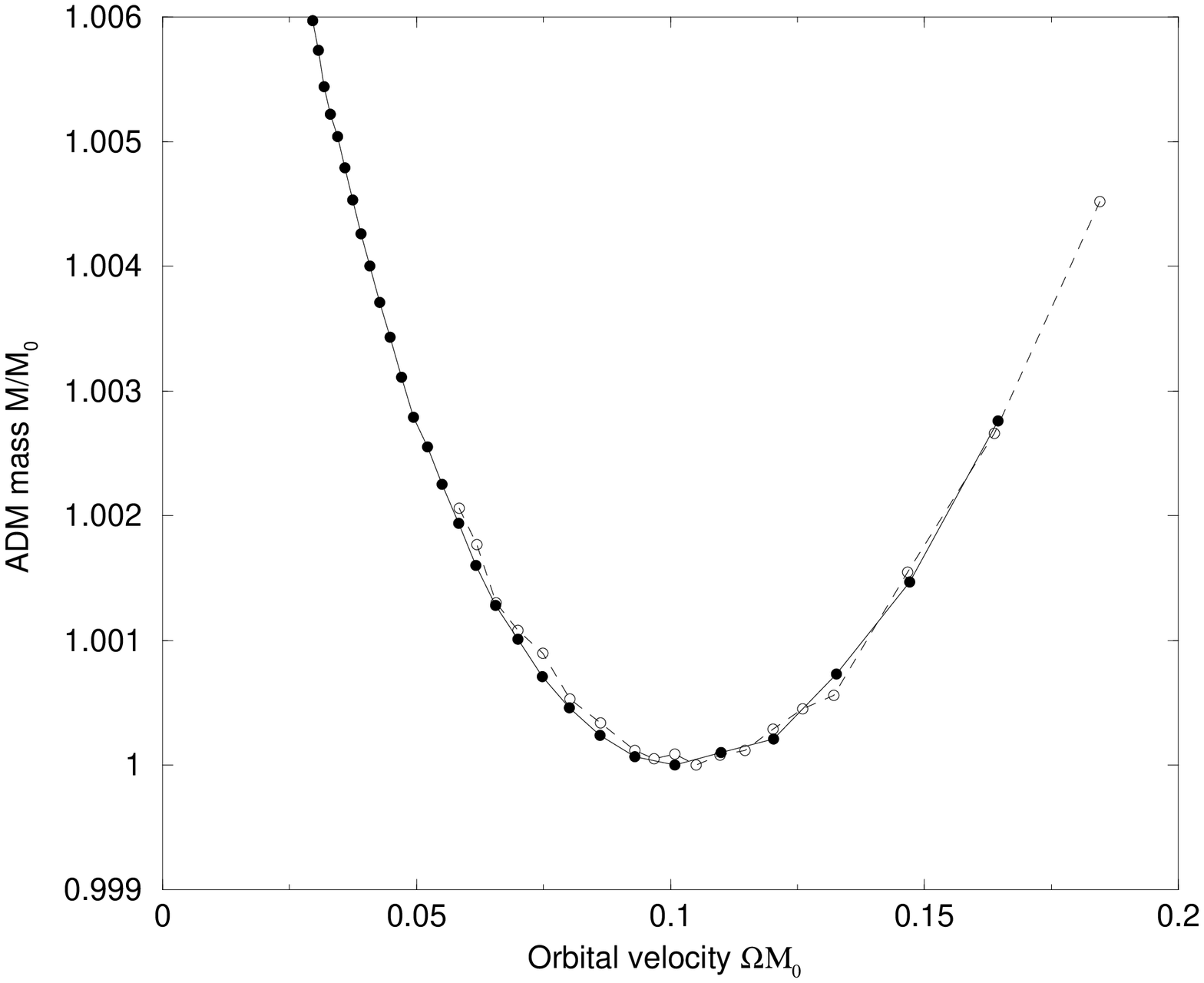}}
\vskip12pt
\caption{\label{f:e_f_ome}
$\bar{M}$ with respect to $\bar{\Omega}$ along a sequence. The filled symbols and solid
line denote the high resolution and the empty symbols and the dashed line the low one.
}
\end{figure}

\begin{figure}
\centerline{\includegraphics[height=9cm,angle=-90]{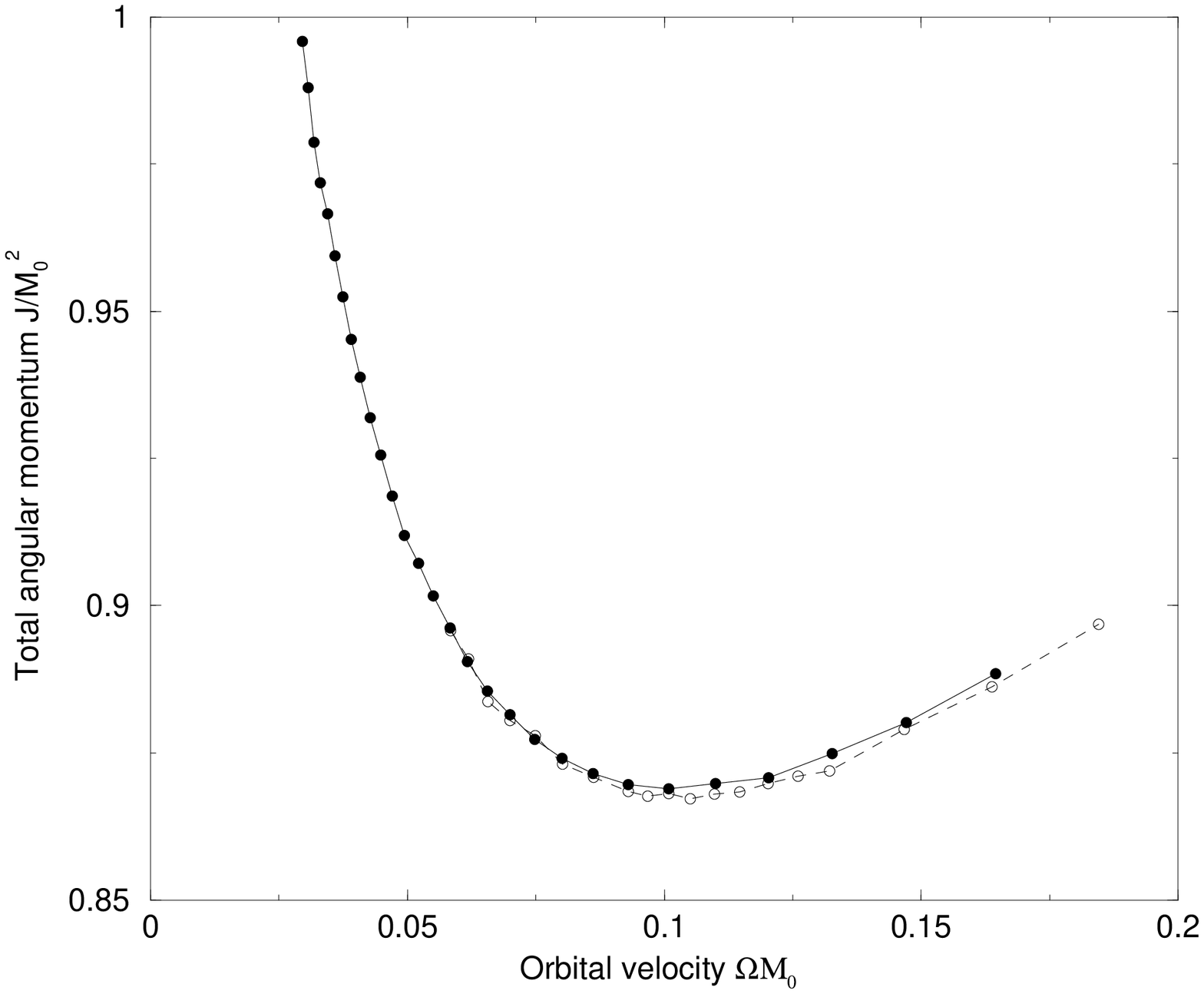}}
\vskip12pt
\caption{\label{f:j_f_ome}
$\bar{J}$ with respect to $\bar{\Omega}$ along a sequence. The filled symbols and solid
line denote the high resolution and the empty symbols and the dashed line the low one.
} 6.60644
\end{figure}

\begin{figure}
\centerline{\includegraphics[height=9cm,angle=-90]{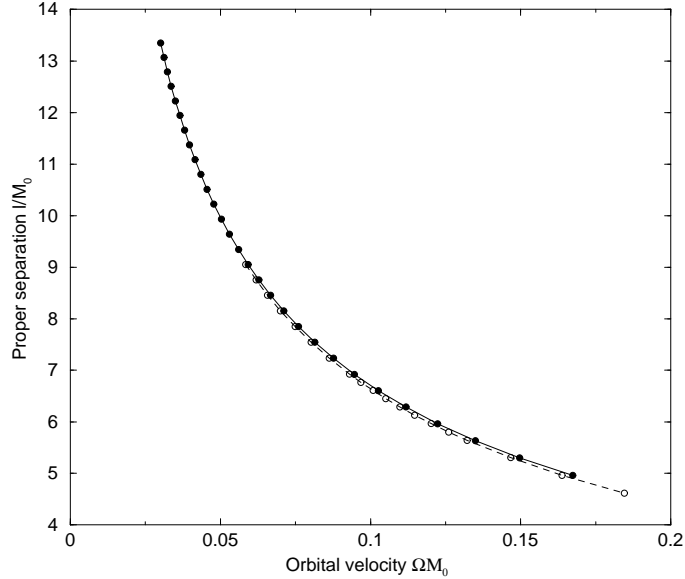}}
\vskip12pt
\caption{\label{f:l_f_ome}
$\bar{l}$ with respect to $\bar{\Omega}$ along a sequence. The filled symbols and solid
line denote the high resolution and the empty symbols and the dashed line the low one.
}
\end{figure}

Figures \ref{f:e_f_j}, \ref{f:e_f_ome}, \ref{f:j_f_ome} and
\ref{f:l_f_ome} show the values
of the dimensionless quantities along a sequence. The calculation has been 
performed with the high and low resolutions
and for values of the parameter $D$ ranging from $40$ to $11$. As previously
mentioned, the sequence exhibits a minimum of $\bar{J}$ and 
$\bar{M}$ as the throats become closer, thereafter interpreted as the signature 
of an innermost stable circular orbit (ISCO) \cite{Baumg01}.
But at this point, we have to be cautious.
Indeed, the relative variation of $\bar{M}$ and $\bar{J}$ along a sequence is 
rather small, and comparable to the precision estimated by means of the 
Smarr formula (see Sec. \ref{s:tests_sequence}).
The exact location of the ISCO being very dependent on those 
small effects, we do not claim to have very precisely determined it.
The following results should be confirmed with more precise calculations.

Another important quantity is the area of the black hole horizons
which relates to the irreducible mass \cite{Chris70} (see also Box 33.4 of 
\cite{MisneTW73}).
As discussed in Sec.~II.B.6 of Paper~I, in our case the
apparent horizons coincide with the two throats.
We therefore define the dimensionless irreducible mass by
\be
\bar{M}_{\rm ir} := \frac{1}{M_0} \left(\sqrt{\frac{A_1}{16\pi}}+\sqrt{\frac{A_2}{16\pi}}
\right) \ ,
\ee
where $A_a$ ($a=1,2$) denotes the area of the throat $a$, calculated according
to the formula
\be
A_a = \oint_{S_a} \Psi^4 d\bar{S}.
\ee

\begin{figure}
\centerline{\includegraphics[height=9cm,angle=-90]{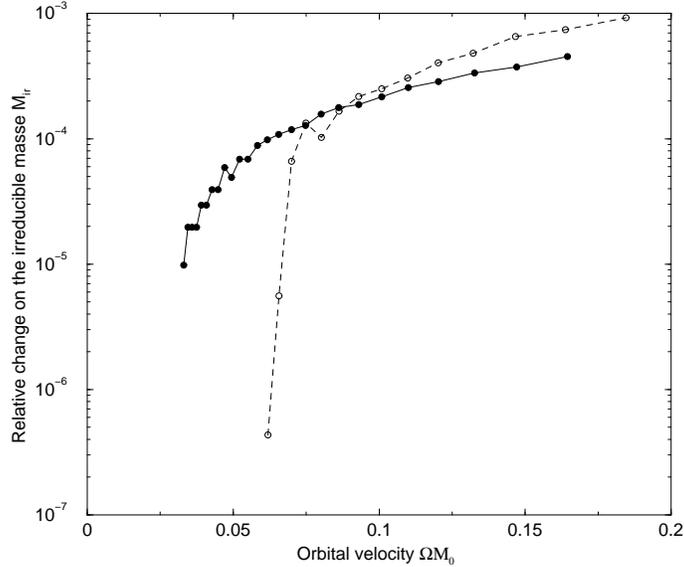}}
\vskip12pt
\caption{\label{f:aire}
Relative change of $\bar{M}_{\rm ir}$ along the sequence, with respect to the orbital 
velocity $\bar{\Omega}$. The filled symbols and the solid line denote the high 
resolution and the empty symbols and the dashed line the low resolution.
}
\end{figure}

Figure \ref{f:aire} shows the relative change of $\bar{M}_{\rm ir}$ along 
the sequence. It
exhibits a slight increase, but its variation is very small. It appears that, 
along the overall sequence, the variation is smaller than $10^{-3}$. The precision 
of our code being of that order, this result is compatible with the fact that 
$\bar{M}_{\rm ir}$ is constant. In other word it shows that 
imposing the condition (\ref{e:def_sequence}) is equivalent to imposing that
the irreducible mass is constant along a sequence.
This constitutes in fact a very good test of our procedure. Indeed Friedman,
Uryu \& Shibata \cite{FriedUS01} have recently established the first law
of binary black hole thermodynamics:
\be
	d M = \Omega d J + \kappa_1 dA_1 + \kappa_2 dA_2  \ , 
\ee
where $\kappa_1$ and $\kappa_2$ are two constants, representing the black
holes surface gravity. For identical black holes ($\kappa_1 = \kappa_2$ and
$dA_1 = dA_2$), the above relation implies
\be
	dM = \Omega dJ \quad \Longleftrightarrow \quad d A_a = 0 \quad (a=1,2) \ . 
\ee
Hence the area of each black hole must be conserved during the evolution. 
In future works, this last criterium could be used to define a sequence, instead
of the relation (\ref{e:def_sequence}).

We choose an average value of the irreducible mass $\bar{M}_{\rm ir} = 1.0173$
and we define then the binding energy of the system at the ISCO by
\be
\left.\bar{E}_{\rm b}\right|_{\rm ISCO} = 1-\bar{M}_{\rm ir},
\ee
the dimensionless total energy being equal to $1$ at the location of the 
turning point.

\begin{figure}
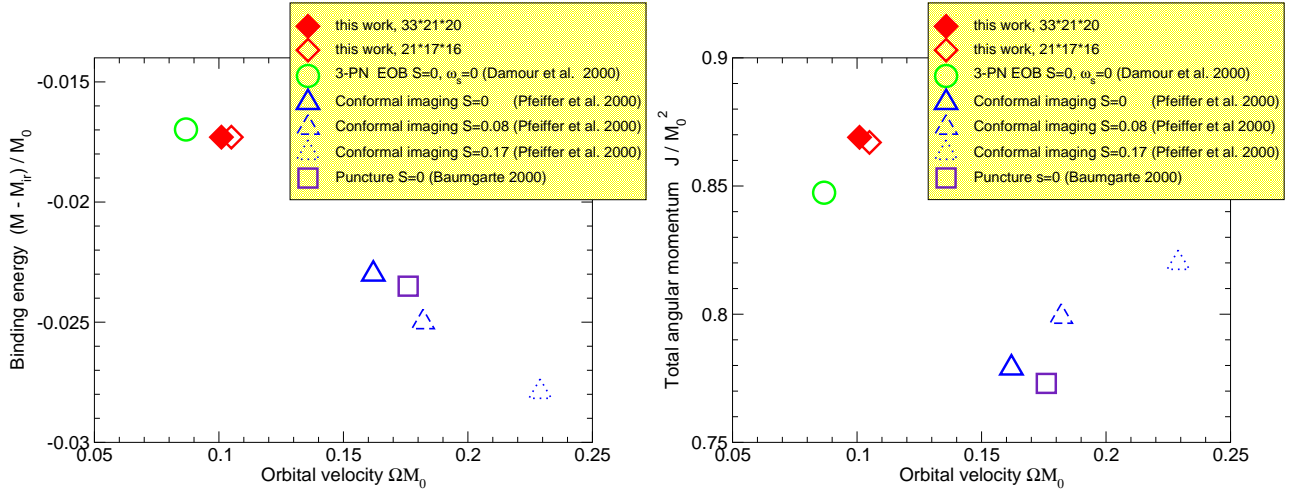

\centerline{\includegraphics[height=6.5cm]{methodes_e.eps}
	    \includegraphics[height=6.5cm]{methodes_j.eps}}
\vskip12pt
\caption{\label{f:methodes} 
Values of $\bar{E}_{\rm b}$ and $\bar{J}$ with respect to $\bar{\Omega}$ at 
the ISCO for different methods, including ours with high and low resolution.
The references to previous studies are as follows: Damour et al.~2000:
\protect\cite{DamouJS00}, 
Pfeiffer et al.~2000: \protect\cite{PfeifTC00} and Baumgarte~2000:
\protect\cite{Baumg00}. 
$S$ denotes the (fixed) spin of the black holes used in various methods.}
\end{figure}

The values of the dimensionless quantities $\bar{\Omega}$, $\bar{J}$, $\bar{E}_{\rm b}$ 
and $\bar{l}$ at the ISCO are 
given in Table~\ref{t:methods} and compared with the results from other
approaches (see \cite{Baumg01} for a review).
{\em 3-PN EOB} stands for the third order post-Newtonian Effective One Body
method for non-spinning black holes \cite{DamouJS00}, with two values 
of the 3-PN parameter $\omega_s$: $\omega_s = 0$ and $\omega_s = -9.34$. 
{\em 3-PN j-method} stands for third order post-Newtonian j-method of 
\cite{DamouJS00}. 
{\em Puncture} denotes the results 
from the puncture method in the case of non-spinning black holes 
\cite{Baumg00} and {\em Conformal imag.} 
the conformal imaging approach with various values of the individual spins 
for rotating black holes \cite{PfeifTC00}. By definition $\bar{M} =1$
at the location of the ISCO for all the methods. The results from the 
different methods are also plotted in Fig. \ref{f:methodes}.

\begin{table}
\caption{Values of dimensionless quantities at the location of the ISCO.
Comparison with other works.}
 \begin{center}
  \begin{tabular}{lcccc}
  Method & $\bar{\Omega}$ & $\bar{J}$ & $\bar{E}_{\rm b}$ & $\bar{l}$
  \\ \hline
3-PN EOB $\omega_s=0$ $S=0$ \cite{DamouJS00} & $0.0868$ & $0.847$ & $-0.0170$ & not given\\
3-PN EOB $\omega_s=-9.34$ $S=0$ \cite{DamouJS00} & $0.0722$ & $0.877$ & $-0.0152$ & not given\\
3-PN j-method  $\omega_s=-9.34$ $S=0$ \cite{DamouJS00}    & $0.0731$ & $0.877$ & $-0.0153$ & not given\\
Puncture $S=0$ \cite{Baumg00}    & $0.176$  & $0.773$ & $-0.0235$ & $4.913$\\
Conformal imag. $S=0$ \cite{PfeifTC00} & $0.162$ & $0.779$ & $-0.0230$ & $5.054$\\
Conformal imag. $S=0.08$ \cite{PfeifTC00} & $0.182$ & $0.799$ & $-0.0250$ & $4.705$\\
Conformal imag. $S=0.17$ \cite{PfeifTC00} & $0.229$ & $0.820$ & $-0.0279$ & $4.040$\\
This work (high res.)  & $0.101$ & $0.869$ & $-0.0173$ & $6.606$\\
This work (low res.)  & $0.105$ & $0.867$ & $-0.0173$ & $6.450$\\
\end{tabular}
 \end{center}
 \label{t:methods}
\end{table}

Figure \ref{f:methodes} shows explicitly that the present results are in much 
better agreement with post-Newtonian calculations than with other numerical works.
Note that the post-Newtonian point plotted on Fig.~\ref{f:methodes} corresponds
to the value $0$ of the (previously ambiguous) 3-PN ``static'' parameter 
$\omega_s$. 
This is indeed the value recently determined by Damour et al. \cite{DamouJS01}. 

But let us point out that it is rather difficult to compare precisely our results
with the other works. The main problem comes from the fact that all those 
methods use individual spins of the black holes as input parameters. In the 
present paper we impose corotation, that is that the throats are spinning 
at the orbital velocity. The only value that can be computed is the total 
angular momentum $J$ and, in general relativity, it cannot be split into
orbital and spins parts in a invariant way.
However, from the results 
of Pfeiffer et al. \cite{PfeifTC00} one can see that increasing the spins of the 
black holes make the values $\bar{\Omega}$, $\bar{J}$ and $-\bar{E}_{\rm b}$ at the 
ISCO
greater. Taking rotation into account in the post-Newtonian methods 
\cite{Damou01} will probably make the orbital velocity and the binding 
energy at the ISCO match even better with our values. 
Work is under progress to compare with
corotating post-Newtonian results \cite{DamouGG01}. 

So, it appears that our results match 
pretty well with post-Newtonian methods. This is the most striking conclusion
from our study. The difference between numerical and post-Newtonian 
results have often been imputed mostly to the conformal flatness 
approximation (see \cite{Baumg01}). The fact that our result, 
{\em using conformal flatness}, is in much better agreement with PN 
calculations than other numerical works, makes us believe that the main 
worry of both conformal imaging and puncture methods lies elsewhere,
possibly in the determination of $\Omega$. Indeed, it is very unlikely that
the orbits and so orbital velocity can be properly computed by solving only for
the four constraint equations. Time should be involved at some level and one 
should take other Einstein equations into account, as we have done here. 

\section{Conclusions}\label{s:conclusion}

The present work should be seen as a first step in trying to give some
new insight to
the binary black holes problem. The basic idea is to extend the numerical 
treatment beyond the resolution of the four constraint equations within a
3-dimensional spacelike surface. This
is achieved by reintroducing time in the problem to deal with a
4-dimensional spacetime.
The orbits are well defined by 
imposing the existence of a helical Killing vector and the orbital velocity 
is found as the only value that equals the ADM and the Komar-like masses,
a requirement which is equivalent to the virial theorem. 
According to us those
are the two most important features of our method. The approximation of 
conformal flatness for the 3-metric
has only been used for simplicity. Sooner or later this problem will have to be 
solved using a general spatial metric and outgoing waves boundary conditions at 
large distances. The use of the inversion isometry to
derive boundary conditions on the throats is also a weak assumption.
In the future, it would be
interesting to change the boundary conditions on the fields in order to 
investigate their influence on the results (see e.g. \cite{Cook01} for 
an alternative choice). 
Besides, changing the boundary conditions on
the shift vector should enable us to describe other states of rotation of 
the black holes, as has been recently proposed by Cook~\cite{Cook01}. 

The numerical schemes are basically the same as those which
have been previously successfully
applied to binary neutron stars configurations \cite{GourgGTMB01}.
They have been extended to solve 
elliptic equations with non-trivial boundary conditions imposed on two throats and 
exact boundary conditions at infinity. Those techniques passed numerous tests and
recover the Schwarzschild and Kerr solutions as well as the Misner-Lindquist one for 
two static black holes \cite{Misne63,Lindq63}.
A technical problem lies in the great number of coefficients 
needed to accurately describe the part of the sources located around 
the companion hole. This effect causes some lack of precision. But we 
can estimate the error it generates by varying the 
number of coefficients, and comparing the results. 
This is what we have done here, using $21\times 17\times 16$
coefficients in each of the 12 domains for the low resolution computations
and $33\times 21\times 20$ coefficients for the high resolution ones. 
The accuracy, estimated from the generalized Smarr formula,
is below $1\%$. 

Another issue is the slight violation of the momentum constraint 
which arises from the necessity to regularize the shift vector. 
We have found that the modification of the shift vector with 
respect to the vector which satisfies the momentum constraint 
(\ref{e:eq_beta}) is below $10^{-3}$, and that the 
error it induces in the momentum constraint equation is
of the order $1\%$. In view of the other approximations
performed in this work, especially the conformal flatness of the 3-metric,
we find this to be very satisfactory. 

In this article, we have defined a sequence of binary black holes
by requiring that the ADM mass decrease is related
to the angular momentum decrease via $dM = \Omega \, dJ$.
This relation is true for the loss due to gravitational radiation, at least
when one considers only the quadrupole formula.
We have then found that the area of the apparent horizons
(irreducible mass) is constant along the sequence, in agreement with 
the first law of binary black holes thermodynamics 
recently derived by Friedman et al. \cite{FriedUS01}. 

The location of the ISCO has been obtained and compared with the results from 
other methods \cite{DamouJS00,Baumg00,PfeifTC00}.
It turns out that our results match the 3-PN methods much better
than previous numerical works. The differences
between numerical studies and 3-PN approximations
have often been explained by the use of the conformal flatness approximation
in the numerical calculations \cite{DamouJS00}.
It seems to us that this is not
the main explanation, for we are using this approximation. 
It certainly arises instead from the way $\Omega$ is determined.

Another natural extension of this work is to use the obtained 
configurations as initial data for binary black holes evolution codes (see 
\cite{Seide99} for a review and Refs. \cite{BrandCGHLL00,AlcubBBLNST01,BakerBCLT01} for
recent results). Initial data files containing the result of the present
work are publically available on the CVS repository of the European
Union Network on Sources of Gravitational Radiation \cite{EUnet}. 
Extraction of the wave-forms from a sequence would also be an interesting
application \cite{DuezBS01,ShibaU01}.

\acknowledgements
This work has benefited from numerous discussions with 
Luc Blanchet, Brandon Carter, Thibault Damour, David Hobill, 
J\'er\^ome Novak and Keisuke Taniguchi. We warmly thank all of them.
The code development and the numerical computations have been performed
on SGI workstations purchased thanks to a special grant from the C.N.R.S.
The public database \cite{EUnet} containing the results is supported by  
the EU Programme 'Improving the Human
Research Potential and the Socio-Economic Knowledge Base' (Research
Training Network Contract HPRN-CT-2000-00137).


\begin{references}

\bibitem{GourgGB01}
E.~Gourgoulhon, P.~Grandcl\'ement and S.~Bonazzola,
submitted to Phys. Rev. D, preprint gr-qc/0106015 [Paper I].

\bibitem{Detwe89}
S.~Detweiler, in {\em Frontiers in Numerical Relativity},
edited by C.R.~Evans, L.S.~Finn, and D.W.~Hobill
(Cambridge University Press, Cambridge, 1989), p.~43.

\bibitem{York79}
J.W.~York, in {\em Sources of Gravitational Radiation}, edited by L.L.~Smarr 
(Cambridge University Press, Cambridge, 1979), p.~83.

\bibitem{MatheMW98}
G.J.~Mathews, P.~Maronetti and J.R.~Wilson,
Phys. Rev. D {\bf 58}, 043003 (1998).

\bibitem{Misne63}
C.W.~Misner, 
Ann. Phys. (N.Y.) {\bf 24}, 102 (1963).

\bibitem{Lindq63}
R.W~Lindquist, 
J. Math. Phys. {\bf 4}, 938 (1963).

\bibitem{KulkaSY83}
A.D.~Kulkarni, L.C.~Shepley, and J.W.~York,
Phys. Lett. {\bf 96A}, 228 (1983). 

\bibitem{Cook91}
G.B.~Cook, 
Phys. Rev. D {\bf 44}, 2983 (1991). 

\bibitem{CookCDKMO93}
G.B.~Cook, M.W.~Choptuik, M.R.~Dubal, S.~Klasky, R.A.~Matzner,
and S.R.~Oliveira, 
Phys. Rev. D {\bf 47}, 1471 (1993). 

\bibitem{Cook94}
G.B.~Cook, 
Phys. Rev. D {\bf 50}, 5025 (1994).
 
\bibitem{PfeifTC00}
H.P.~Pfeiffer, S.A.~Teukolsky, and G.B.~Cook, 
Phys. Rev. D {\bf 62}, 104018 (2000).

\bibitem{DieneJKN00}
P.~Diener, N.~Jansen, A.~Khokhlov, and I.~Novikov,
Class. Quantum Grav. {\bf 17}, 435 (2000).

\bibitem{GourgB94}
E.~Gourgoulhon and  S.~Bonazzola, 
Class. Quantum Grav. {\bf 11}, 443 (1994). 

\bibitem{GourgGTMB01}
E.~Gourgoulhon, P.~Grandcl\'ement, K.~Taniguchi, J.-A.~Marck, 
and S.~Bonazzola,
Phys. Rev. D  {\bf 63}, 064029 (2001). 

\bibitem{CanutHQZ88}
C.~Canuto, M.Y.~Hussaini, A.~Quarteroni and T.A.~Zang
{\em Spectral Methods in Fluid Dynamics} (Spinger-Verlag,Berlin,1988)

\bibitem{BonazGM98}
S.~Bonazzola, E.~Gourgoulhon and J.-A.~Marck,
Phys. Rev. D {\bf 58}, 104020 (1998).

\bibitem{BonazGM99}
S.~Bonazzola, E.~Gourgoulhon and J.-A.~Marck,
J. Comput. Appl. Math. {\bf 109}, 433 (1999).

\bibitem{GrandBGM01}
P.~Grandcl\'ement, S.~Bonazzola, E.~Gourgoulhon and J.-A.~Marck,
J. Comp. Phys. {\bf 170}, 231 (2001). 

\bibitem{JanseDKN01}
N.~Jansen, P.~Diener, A.~Khokhlov, and I.~Novikov,
preprint: gr-qc/0103029.

\bibitem{OoharN97}
K.~Oohara and T.~Nakamura, in {\em Relativistic Gravitation and
Gravitational Radiation}, edited by J.-A.~Marck and J.-P.~Lasota
(Cambridge University Press, Cambridge, 1997), p.~309.

\bibitem{OoharNS87}
K.~Oohara, T.~Nakamura and M.~Shibata,
Prog. Theor. Phys. Suppl. {\bf 128}, 183 (1987).

\bibitem{Cook01}
G.B.~Cook, 
preprint gr-qc/0108076. 

\bibitem{GaratP00}
A.~Garat and R.H.~Price,
Phys. Rev. D {\bf 61}, 124011 (2000).

\bibitem{Giuli98}
D.~Giulini,
in {\em Black Holes: Theory and Observation},
edited by F.W.~Hehl, C.~Kiefer and R.J.K.~Metzler (Spinger-Verlag, Berlin,
1998), p.~224.

\bibitem{AndraP97}
Z.~Andrade and R.~Price,
Phys. Rev. D {\bf 56}, 6336 (1997).

\bibitem{ShapiT83}
S.L.~Shapiro and S.A.~Teukolsky,
{\em Black Holes, White Dwarfs and Neutron Stars},
(J.~Wiley and Sons, New-York,1983).

\bibitem{BaumgCSST98}
T.W. Baumgarte, G.B. Cook, M.A. Scheel, S.L. Shapiro, and S.A. Teukolsky,
\newblock Phys. Rev. D {\bf 57}, 7299 (1998).

\bibitem{UryuSE00} 
K. Uryu, M.~Shibata, and Y. Eriguchi, 
\newblock Phys. Rev. D {\bf 62}, 104015 (2000). 

\bibitem{Baumg01}
T.W.~Baumgarte,
in {\em Astrophysical Sources of Gravitational Radiation},
edited by J.M.~Centrella, AIP press, in press,
preprint gr-qc/0101045.

\bibitem{Chris70}
C.~Christodoulou,
Phys. Rev. Lett. {\bf 25}, 1596 (1970).

\bibitem{MisneTW73}
C.W.~Misner, K.S.~Thorne, and J.A.~Wheeler, 
{\em Gravitation} (Freeman, New York, 1973).

\bibitem{FriedUS01}
J.L.~Friedman, K.~Uryu, and M.~Shibata, 
preprint gr-qc/0108070. 

\bibitem{DamouJS00}
T.~Damour, P.~Jaranowski and G.~Sch\"afer
Phys. Rev. D {\bf 62}, 084011 (2000).

\bibitem{Baumg00}
T.W.~Baumgarte
Phys. Rev. D {\bf 62}, 024018 (2000).

\bibitem{DamouJS01}
T.~Damour, P.~Jaranowski and G.~Sch\"afer,
Phys. Lett. B {\bf 513}, 147 (2001).

\bibitem{Damou01}
T.~Damour,
Phys. Rev. D, in press (preprint: gr-qc/0103018).

\bibitem{DamouGG01}
T.~Damour, E.~Gourgoulhon, and P.~Grandcl\'ement,
in preparation.

\bibitem{Seide99}
E.~Seidel,
in {\em Black Holes and Gravitational Waves -- New Eyes in
the 21st Century, Proc. of the 9th Yukawa International Seminar,
Kyoto 1999}, edited by T.~Nakamura and H.~Kodama,
Prog. Theor. Phys. Suppl. {\bf 136}, 87 (1999).

\bibitem{BrandCGHLL00}
S.~Brandt, R.~Correll, R.~G\'omez, M.~Huq, P.~Laguna, L.~Lehner, P.~Marronetti,
R.A.~Matzner, D.~Neilsen, J.~Pullin, E.~Schnetter, D.~Shoemaker,
and J.~Winicour, Phys. Rev. Lett. {\bf 85}, 5496 (2000). 

\bibitem{AlcubBBLNST01}
M.~Alcubierre, W.~Benger, B.~Br\"ugmann, G.~Lanfermann, L.~Nerger, E.~Seidel, 
and R.~Takahashi, preprint gr-qc/0012079. 

\bibitem{BakerBCLT01}
J. Baker, B. Br\"ugmann, M. Campanelli, C. O. Lousto, and R. Takahashi,
Phys. Rev. Lett. {\bf 87}, 121103 (2001). 

\bibitem{EUnet}
http://www.eu-network.org/Projects/InitialData.html

\bibitem{DuezBS01}
M.D.~Duez, T.W.~Baumgarte and S.L.~Shapiro
Phys. Rev. D {\bf 63}, 084030 (2001).

\bibitem{ShibaU01}
M.~Shibata and K.~Uryu,
Phys. Rev. D, in press (preprint: gr-qc/0109026).

\end{references}
\end{document}